\documentclass{article}
\usepackage[utf8]{inputenc}
\usepackage[margin=1in]{geometry}
\usepackage{amsmath}
\usepackage[hidelinks]{hyperref}
\usepackage{cleveref}
\usepackage{apacite}
\usepackage{natbib}
\usepackage{authblk}
\usepackage{graphicx}
\usepackage{subcaption}
\graphicspath{{figures/}}
\usepackage{multirow}
\makeatletter
\def\blfootnote{\xdef\@thefnmark{*}\@footnotetext}
\makeatother
\usepackage{setspace}
\title{QuakeFlow: A Scalable Machine-learning-based Earthquake Monitoring Workflow with Cloud Computing}
\author[1,2]{Weiqiang Zhu}
\author[3]{Alvin Brian Hou}
\author[3]{Robert Yang}
\author[4]{Avoy Datta}
\author[2]{S. Mostafa Mousavi}
\author[2]{William L. Ellsworth}
\author[2*]{Gregory C. Beroza}

\affil[1]{\small Seismological Laboratory, California Institute of Technology, Pasadena, CA 91125}
\affil[2]{\small Department of Geophysics, Stanford University, Stanford, CA, 94305}
\affil[3]{\small Computer Science Department, Stanford University, Stanford, CA, 94305}
\affil[4]{\small Electrical Engineering Department, Stanford University, Stanford, CA, 94305}
\date{}

\begin{document}

\maketitle
\blfootnote{Corresponding author: beroza@stanford.edu}

\begin{abstract}
Earthquake monitoring workflows are designed to detect earthquake signals and to determine source characteristics from continuous waveform data. Recent developments in deep learning seismology have been used to improve tasks within earthquake monitoring workflows that allow the fast and accurate detection of up to orders of magnitude more small events than are present in conventional catalogs. To facilitate the application of machine-learning algorithms to large-volume seismic records at scale, we developed a cloud-based earthquake monitoring workflow, QuakeFlow, that applies multiple processing steps to generate earthquake catalogs from raw seismic data. QuakeFlow uses a deep learning model, PhaseNet, for picking P/S phases and a machine learning model, GaMMA, for phase association with approximate earthquake location and magnitude. Each component in QuakeFlow is containerized, allowing straightforward updates to the pipeline with new deep learning/machine learning models, as well as the ability to add new components, such as earthquake relocation algorithms. We built QuakeFlow in Kubernetes to make it auto-scale for large datasets and to make it easy to deploy on cloud platforms, which enables large-scale parallel processing. We used QuakeFlow to process three years of continuous archived data from Puerto Rico within a few hours, and found more than a factor of ten more events that occurred on much the same structures as previously known seismicity. We applied Quakeflow to monitoring frequent earthquakes in Hawaii and found over an order of magnitude more events than are in the standard catalog, including many events that illuminate the deep structure of the magmatic system. We also added Kafka and Spark streaming to deliver real-time earthquake monitoring results. QuakeFlow is an effective and efficient approach both for improving realtime earthquake monitoring and for mining archived seismic data sets.
\end{abstract}


\section{Introduction}

Continuous seismic waveforms are recorded across seismic networks and processed by earthquake monitoring workflows to detect, locate, and characterize seismic events. The resulting earthquake catalogs illuminate the 3D geometry of seismically active structures and reveal the spatio-temporal evolution of seismicity. A comprehensive earthquake catalog provides crucial information for understanding complex earthquake sequences and quantifying earthquake hazard. The increasing number of dense seismic networks and the rapidly accumulating amount of seismic waveform data over time pose a challenge for mining seismic datasets to realize the full benefit of more extensive instrumentation. The rapid progress of deep learning algorithms when coupled with cloud computing provides a promising pathway to address the big data challenge in earthquake monitoring to generate comprehensive and accurate earthquake catalogs.

Deep learning seismology \citep{Mousavi2022DL} has dramatically improved earthquake monitoring performance particularly in earthquake detection and phase picking \citep{perol2018convolutional, ross2018generalized, zhu2019phasenet, mousavi2020earthquake}. Earthquake monitoring workflows using deep-learning-based phase pickers have been applied to studying, dense earthquake sequences \citep{liu2020rapid, ross20203d, tan2021machine}, induced seismicity \citep{park2020machine, wang2020injection, chai2020using, park2022Oklahoma, zhou2021machine}, marine seismicity \citep{gong2022microseismicity, jiang2022detailed}, and magmatic systems \citep{retailleau2022automatic}. These studies have demonstrated that deep learning can: detect up to orders of magnitude more small earthquakes than conventional algorithms, provide a more complete accounting of seismicity, and enable new insight into earthquake behavior. Deep-learning-based earthquake monitoring workflows are now feasible for adoption in operational earthquake monitoring systems and that transition is proceeding apace \citep{yeck2021leveraging, walter2021easyquake, zhang2022loc, retailleau2022wrapper, shi2022malmi}. 

Massive amounts of continuous seismic data have accumulated over the last several decades. As of April, 2022 the archived data volume at the Incorporated Research Institutions for Seismology (IRIS) has reached about 800 TiB\footnote{\url{http://ds.iris.edu/data/distribution/}}. There are many times this much data from local and regional earthquake monitoring networks that are stored elsewhere.  With the advent of distributed acoustic sensing (DAS) for earthquake monitoring, the rate of data accumulation is poised to accelerate dramatically \citep{zhan2020distributed,lindsey2021fiber}. The current data access model, in which individual users download all data of interest for local processing, would seem to have a limited future. Cloud computing has the potential to address the coupled big data challenges posed by the need to access, and the need to process, massive seismic datasets. A recent survey of international seismological data users \citep{quinteros2021exploring} expressed the need for cloud-ready seismic processing software as part of a future vision for cloud-based seismic waveform data repositories. Cloud computing can significantly decrease the wall time required for data processing by using thousands of computational nodes provided by cloud computing platforms, such as AWS (Amazon Web Services), GCP (Google Cloud Platform), and Microsoft Azure. 
Cloud-native computing technologies enable building and deploying scalable applications in the cloud.
For example, Docker\footnote{\url{https://www.docker.com/}} is widely used to package an application and its dependencies in a virtual container that can flexibly run on different computing platforms.
Kubernetes\footnote{\url{https://kubernetes.io/}} is a container orchestration system that deploys, maintains and scales applications based on computational workloads such as CPU, memory, and storage.
Kubeflow\footnote{\url{https://www.kubeflow.org/}} is a cloud-native framework to run machine learning pipelines on Kubernetes clusters. 
Kafka\footnote{\url{https://kafka.apache.org/}} is a distributed message streaming service that allows writing, reading, and storing data streams.
Spark streaming\footnote{\url{https://spark.apache.org/}} is a scalable stream processing engine that performs analytics on streaming data from messaging services such as Kafka.
FastAPI\footnote{\url{https://fastapi.tiangolo.com/}} is a web framework for developing RESTful APIs, where REST stands for Representational State Transfer, a software architectural style that describes a uniform interface in a client-server architecture, and API stands for Application Programming Interface.
These rapidly developing cloud-native softwares, particularly when coupled with cloud-based datasets, such as the SCEDC AWS public dataset\footnote{\url{https://scedc.caltech.edu/data/cloud.html}}, make cloud-based processing particularly efficient.

In this study, we combined two machine learning algorithms for earthquake detection with cloud computing for parallel processing, to build an earthquake monitoring workflow that we call ``QuakeFlow".  QuakeFlow can be applied to either mining massive archived datasets or to processing real-time streamed waveforms. For the initial application of QuakeFlow we used the deep neural network model, PhaseNet \citep{zhu2019phasenet}, to pick P- and S-phases and the Gaussian mixture model, GaMMA \citep{zhu2022earthquake}, to associate picks and estimate approximate earthquake locations and magnitudes. We added Kafka and Spark Streaming services to support real-time earthquake monitoring, and deployed the QuakeFlow system in Kubernetes, making it platform-independent and scalable to thousands of cloud computing nodes for large-scale seismic data mining. QuakeFlow is set up to combine state-of-the-art machine learning models and cloud computing techniques for earthquake monitoring.

\section{Earthquake monitoring workflow}

\begin{figure}[htp]
    \centering
    \begin{subfigure}{\textwidth}
    \includegraphics[width=\linewidth]{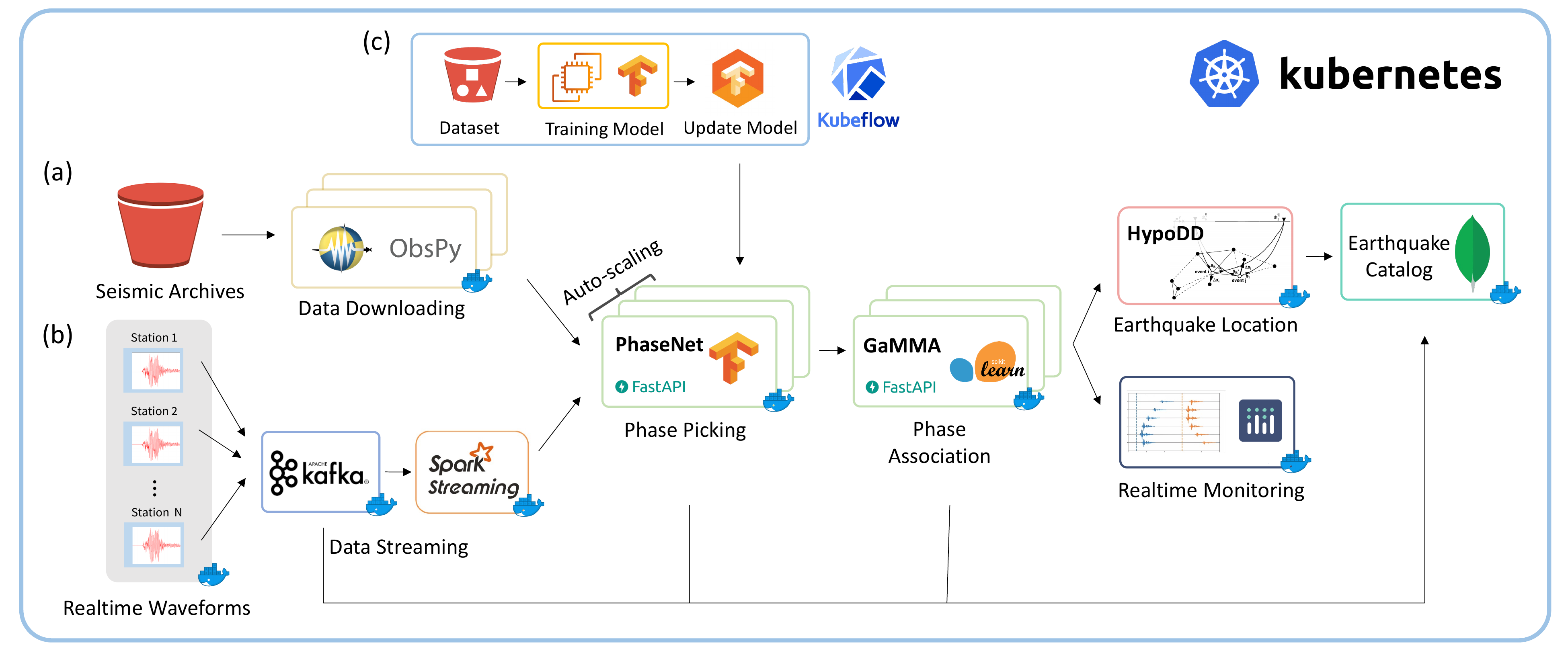}
    \end{subfigure}
    \caption{QuakeFlow diagram: (a) batch prediction for mining seismic archives; (b) stream prediction for real-time monitoring; (c) training pipeline for updating deep learning models. Each component of QuakeFlow is containerized using Docker and orchestrated using Kubernetes with auto-scaling. We implement both PhaseNet and GaMMA models and support both large-scale parallel processing of archived seismic waveforms and stream processing of realtime seismic waveforms. }
    \label{fig:system}
\end{figure}

An earthquake monitoring workflow consists of a sequence of tasks: phase detection/picking, association, location, and characterization, to detect earthquake signals and estimate source parameters. In this proof-of-concept study, we focused on two of these tasks: phase picking and phase association, which have been significantly improved by machine learning models. Models such as these are not yet widely used by operational earthquake monitoring systems. We built the QuakeFlow system to explore the potential of machine learning algorithms and cloud computing infrastructure for efficient earthquake monitoring. \Cref{fig:system} shows an overview of QuakeFlow. Two key modules used by QuakeFlow are a deep learning model, PhaseNet, for picking P- and S-phase arrival times and a Gaussian mixture model, GaMMA, for associating phases and estimating approximate earthquake locations and magnitudes. Both models are containerized (meaning that the software is packaged with all necessary dependencies) using Docker, and deployed in Kubernetes. These models can be flexibly replaced by other machine learning models, such as GPD \citep{ross2018generalized} or EQTransformer \citep{mousavi2020earthquake} for phase picking and REAL \citep{zhang2019rapid} or PhaseLink \citep{ross2019phaselink} for phase association, once containerized. We enable auto-scaling for parallel processing using Kubernetes, and built a batch prediction pipeline that can mine archived datasets in an embarrassingly parallel manner, which means that the dataset can be distributed into a number of parallel processes without communication between them, using KubeFlow (\Cref{fig:system}a). We used ObsPy \citep{beyreuther2010obspy} to retrieve continuous seismic waveforms from seismic data centers. These waveforms are processed using PhaseNet and GaMMA in parallel, and then relocated using HypoDD, a joint earthquake relocation algorithm \citep{waldhauser2001hypodd}. In addition to the default batch processing for mining archived datasets, we added Kafka and Spark Streaming services to support stream processing for real-time monitoring (\Cref{fig:system}b). We received real-time seismic waveforms from seismic networks using the SeedLink API\footnote{\url{http://ds.iris.edu/ds/nodes/dmc/services/seedlink/}} and continuously pushed these waveforms to the Kafka messaging service. We used the Spark Streaming service to build an ETL (extract, transform, load) pipeline, which applies a sequence of MapReduce transformations, such as windowing, grouping, filtering, and aggregation. The pre-processed data is then sent to PhaseNet and GaMMA APIs that are exposed as RESTful services using FastAPI. The detected earthquakes with approximate locations and magnitudes are saved to an earthquake catalog using the MongoDB database\footnote{\url{https://www.mongodb.com/}} and broadcast through Kafka to a web app to display real-time waveform and earthquake information. We also built a training and inference pipeline based on the Kubeflow framework for training and updating machine learning models on the cloud (\Cref{fig:system}c). We note that QuakeFlow is not limited to its current implementation. New models and more components such as earthquake location algorithms \citep{klein2002user,lomax2000probabilistic,smith2022hyposvi} can be containerized and added to this earthquake monitoring workflow.

\subsection{Machine learning models}

\begin{figure}[htp]
    \centering
    \begin{subfigure}{0.33\textwidth}
    \includegraphics[width=\textwidth]{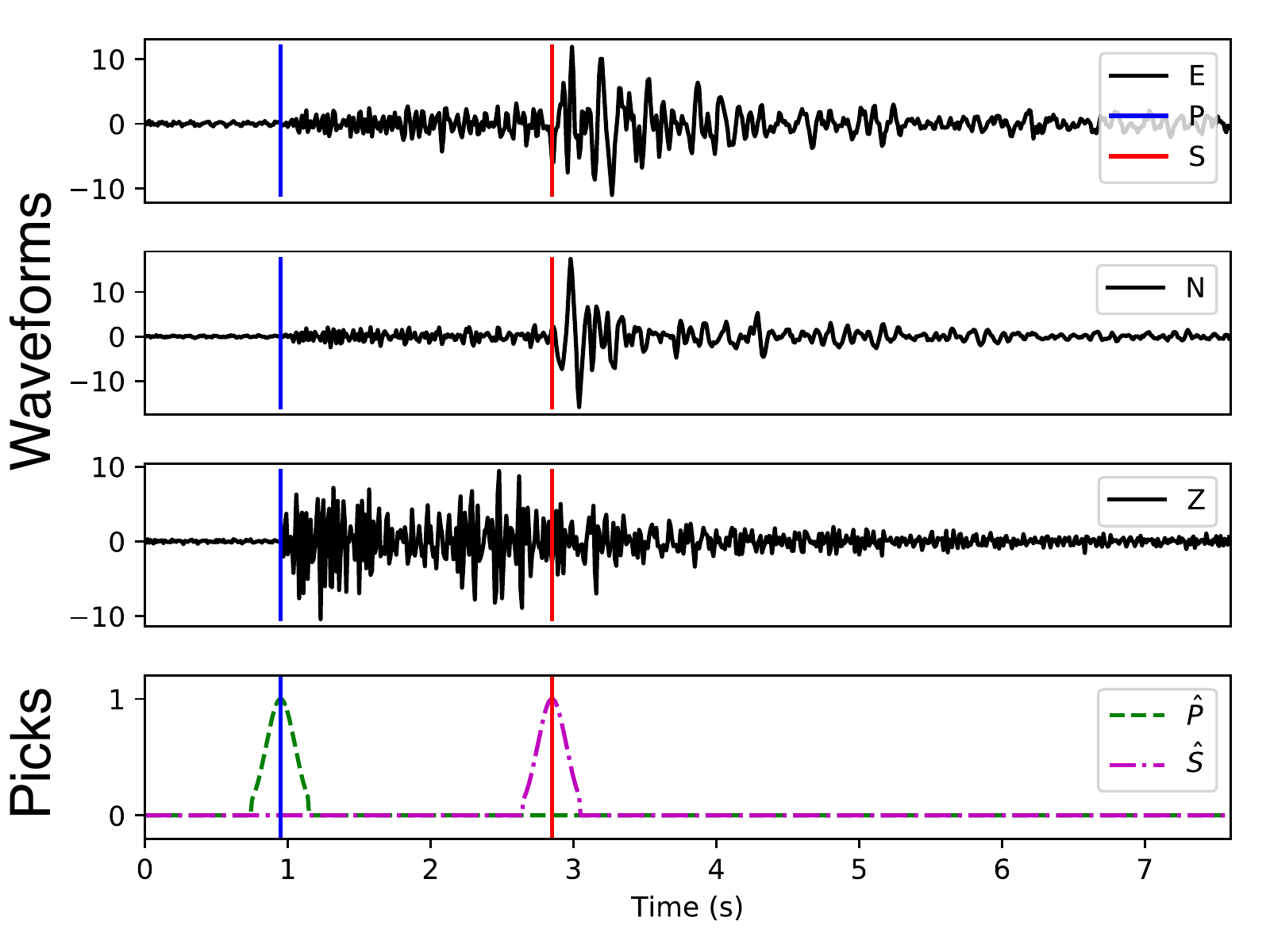}
    \caption{}
    \end{subfigure}
    \begin{subfigure}{0.6\textwidth}
    \includegraphics[width=\textwidth]{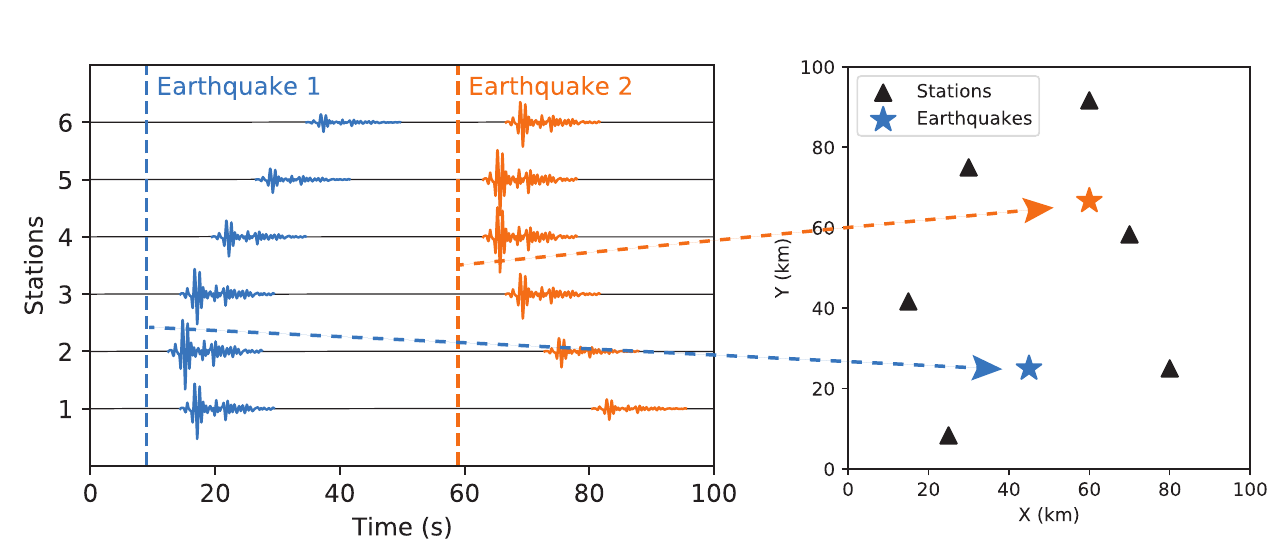}
    \caption{}
    \end{subfigure}
    \begin{subfigure}{0.55\textwidth}
    \includegraphics[width=\textwidth]{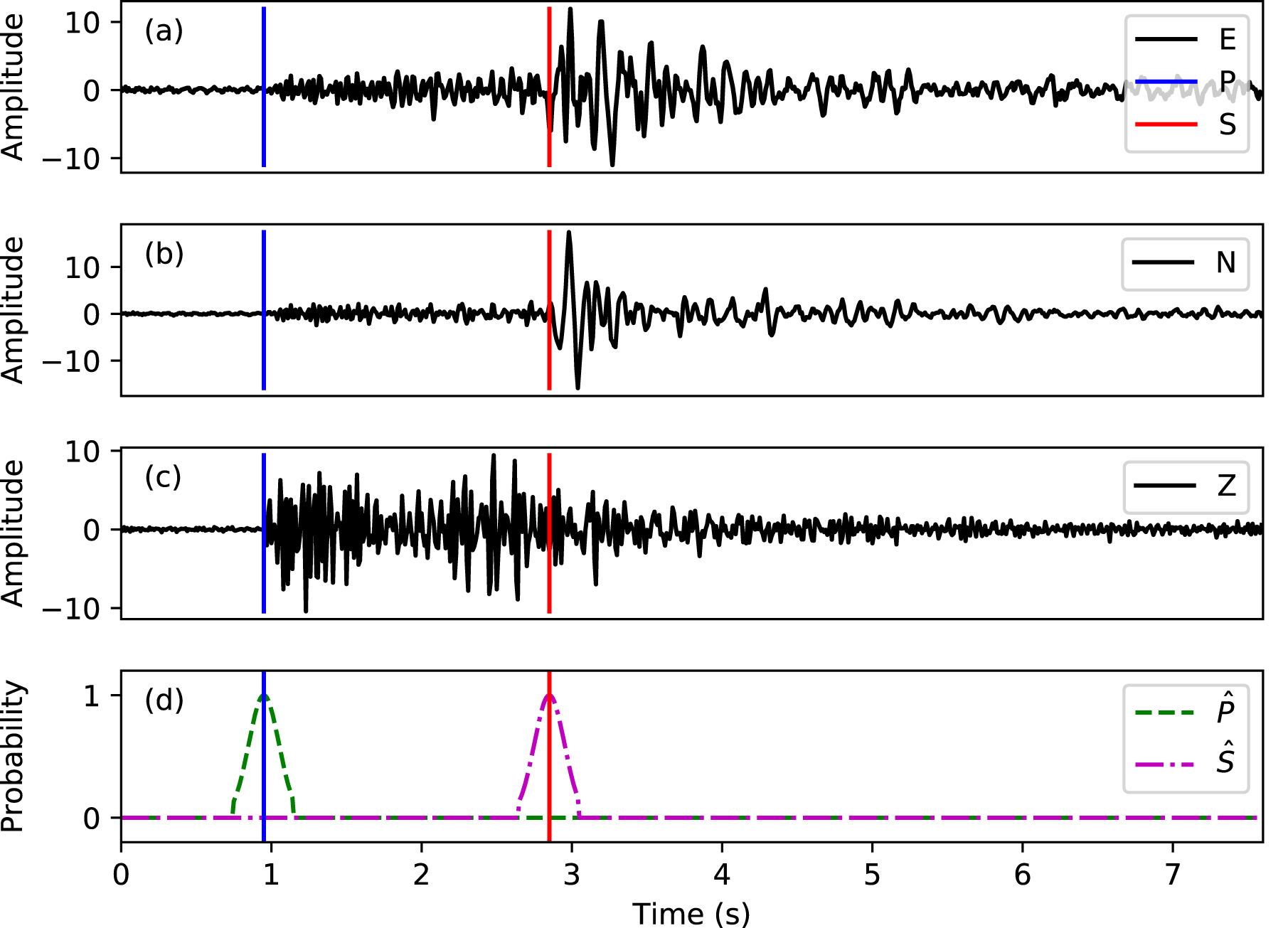}
    \caption{}
    \end{subfigure}
    \begin{subfigure}{0.37\textwidth}
    \includegraphics[width=\textwidth]{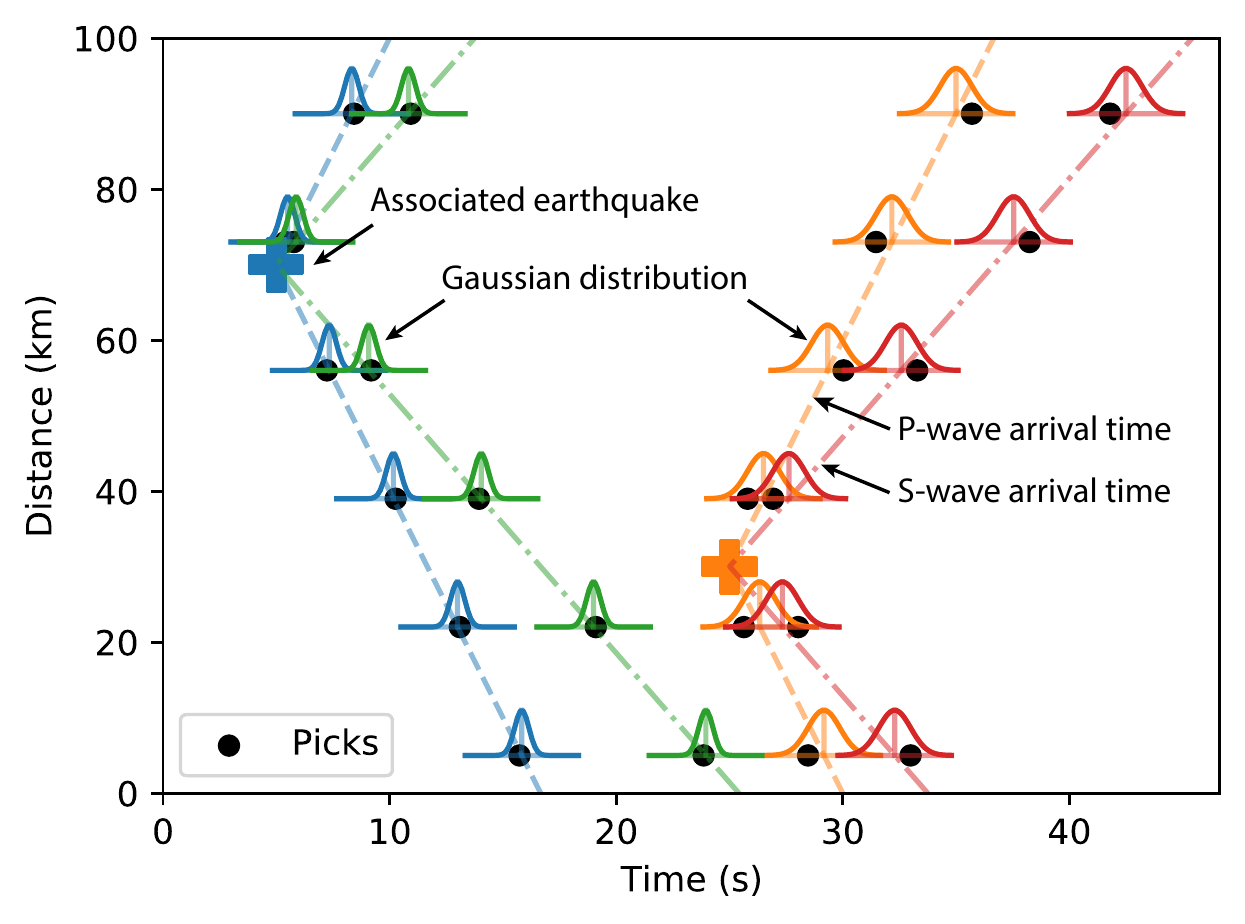}
    \caption{}
    \end{subfigure}
    \caption{Upper two panels depict two tasks in the earthquake monitoring workflow: (a) Picking \textit{P} and \textit{S} phases from waveforms for each seismic station. (b) Associating \textit{P} and \textit{S} phases picked across multiple stations and determining earthquake locations and magnitudes. Lower two panels show (c) the PhaseNet model used for phase picking and (d) the GaMMA model used for phase association. }
    \label{fig:method_quakeflow}
\end{figure}

Phase picking and phase association are two key tasks for earthquake monitoring (\Cref{fig:method_quakeflow}). New AI-based phase pickers can detect and pick the arrival times of seismic phases, i.e., \textit{P}-phase and \textit{S}-phase (\Cref{fig:method_quakeflow}a) at each seismic station; phase association assembles these phase picks into associated arrivals from causative earthquakes, or unassociated picks that are from either insufficiently well recorded earthquakes or from non-earthquake noise sources (\Cref{fig:method_quakeflow}b). In QuakeFlow, we replace conventional phase picking and phase association algorithms with a deep neural network model, PhaseNet (\Cref{fig:method_quakeflow}c), and a Gaussian mixture model, GaMMA (\Cref{fig:method_quakeflow}d). The two models together extract time and amplitude information of \textit{P} and \textit{S} phases, and detect earthquakes with approximate earthquake locations and magnitudes. 

\paragraph{PhaseNet} We used the pre-trained PhaseNet model \citep{zhu2019phasenet} to pick the arrival times of \textit{P} and \textit{S} phases from continuous seismic waveforms (\Cref{fig:method_quakeflow}c). PhaseNet is a convolutional neural network (CNN) model that effectively predicts two Gaussian-shaped characteristic functions for \textit{P} and \textit{S} phases, from which we can extract accurate arrival times. By training on more than 700k examples labeled by human analysts, PhaseNet achieves a much better picking performance than conventional algorithms - typically detecting an order of magnitude more S-picks with high precision and low bias.

\paragraph{GaMMA} We use the unsupervised GaMMA algorithm \citep{zhu2022earthquake}, which stands for \textbf{Ga}ussian \textbf{M}ixture \textbf{M}odel \textbf{A}ssociation, to associate picked phase arrivals from multiple stations across a seismic network. GaMMA treats earthquake phase association as an unsupervised clustering problem in a probabilistic framework, where phases are clustered following an approximately hyperbolic moveout of phase travel times (\Cref{fig:method_quakeflow}d). The probabilistic framework flexibly considers multiple types of phase information, including arrival time, amplitude, phase type, and quality score, to associate phases from a dense earthquake sequence effectively.

\subsection{Cloud computing}

\paragraph{Data streaming} Apache Kafka is a fault-tolerant, highly scalable, distributed messaging system for streaming applications \citep{kreps2011kafka}. In QuakeFlow, Kafka acts as the central hub for real-time streaming of waveform data and model prediction results. Our Kafka settings are as follows: there are three pre-defined Kafka topics: \texttt{waveform\_raw}, \texttt{phasenet\_picks} and \texttt{gamma\_events}. The monitoring stations continuously send fragments of seismic waveforms to topic \texttt{waveform\_raw}. We then use the scalable, fault-tolerant Spark Streaming processing system, which supports both batch and streaming workloads \citep{zaharia2013discretized}, as an ETL pipeline to apply data transformations and pre-processing to the streaming data. Spark Streaming supports operations to aggregate streaming data over a sliding window. We group the streaming data in a specified window size (e.g., 30 seconds) using a sequence of MapReduce operations to prepare a structured data format for subsequent processing of PhaseNet and GaMMA. The outputs of these machine learning models are broadcast to the \texttt{phasenet\_picks} and \texttt{gamma\_events} topics, respectively. Finally, the earthquake detection results can be saved and visualized by subscribing to the corresponding Kafka topics.

\begin{figure}[htp]
    \centering
    \begin{subfigure}{0.48\textwidth}
    \includegraphics[width=\textwidth]{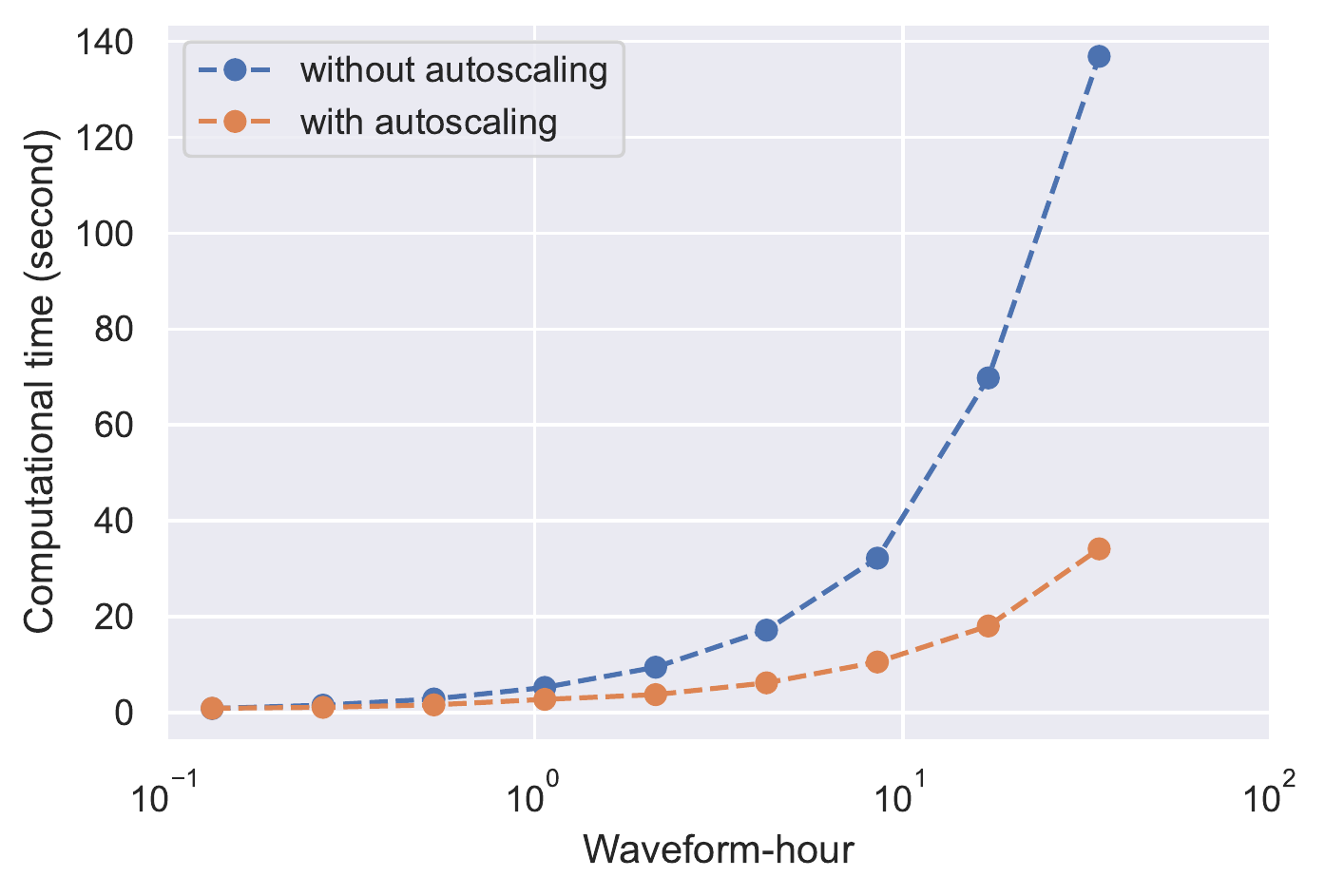}
    \caption{}
    \end{subfigure}
    \begin{subfigure}{0.48\textwidth}
    \includegraphics[width=\textwidth]{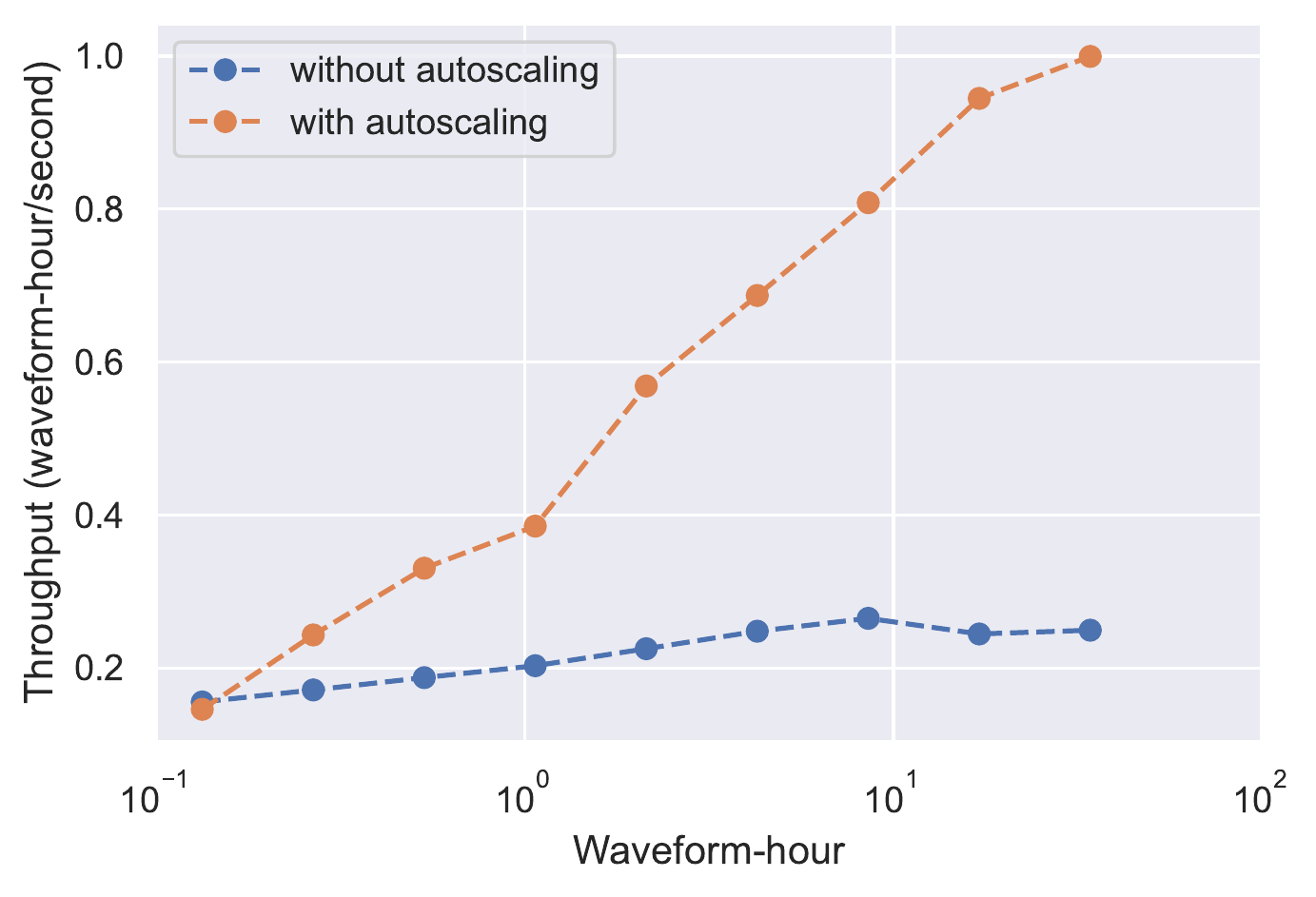}
    \caption{}
    \end{subfigure}
    \caption{Speedup by auto-scaling: (a) computational time; (b) data throughput.}
    \label{fig:autoscaling}
\end{figure}

\paragraph{Auto-scaling} We deployed QuakeFlow in the Kubernetes system, making it platform-independent and applicable to both on-premise servers and any cloud-platforms with Kubernetes services. Kubernetes automatically orchestrates different components of QuakeFlow to make it run on the cloud efficiently. We used both the horizontal pod auto-scaling provided by Kubernetes, as well as the node auto-provision provided by cloud-platforms, such as Google Cloud Platform (GCP), to  match computational resources automatically with computational load. We carried out a simple pressure test on GCP using a maximum of 8 computational nodes of machine type ``n2-standard-2"\footnote{\url{https://cloud.google.com/compute/docs/machine-types}} (2 vCPU and 8GB of memory) to evaluate the speedup using auto-scaling when processing a large data volume. \Cref{fig:autoscaling}a shows that the computational time with auto-scaling is significantly reduced compared to the computational time without it. Data throughput, i.e. the number of waveform-hours processed per second, linearly increases with the data volume when auto-scaling is enabled (\Cref{fig:autoscaling}b). This means that we can apply auto-scaling for embarrassingly parallel large-scale seismic data mining.

\section{Applications}

We applied QuakeFlow to study two cases: tectonic earthquakes in Puerto Rico and volcanic earthquakes in Hawaii. 

\subsection{Earthquake detection in Puerto Rico}

An earthquake sequence in Puerto Rico started on December 28, 2019 and continued through 2021. The largest earthquake (M 6.4) to date occurred on January 7, 2020, causing many injuries and widespread damage \citep{vanacore2022preface}. The sequence was rich in seismicity \citep{vivcivc20222019} and involved both strike slip and normal faulting \citep{ten2022mature} with much of the activity on the Punta Montalva and Guayanilla Canyon Faults. Deformation in this area is transtensional and diffuse, with unmapped faults capable of generating moderate-to-large earthquakes \citep{viltres2022transtensional}. 

\begin{figure}
    \centering
    \begin{subfigure}{0.48\textwidth}
    \includegraphics[width=\textwidth]{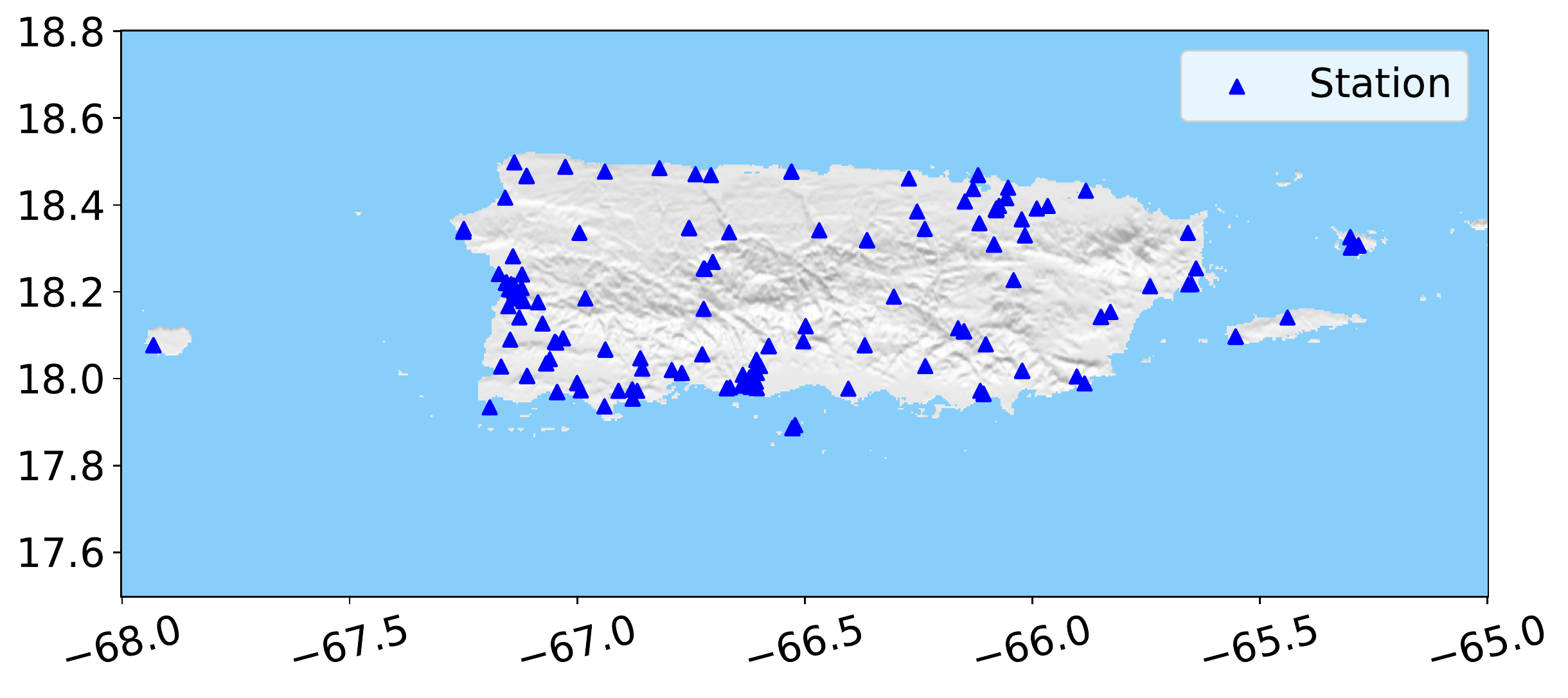}
    \caption{}
    \end{subfigure}
    \begin{subfigure}{0.48\textwidth}
    \includegraphics[width=\textwidth]{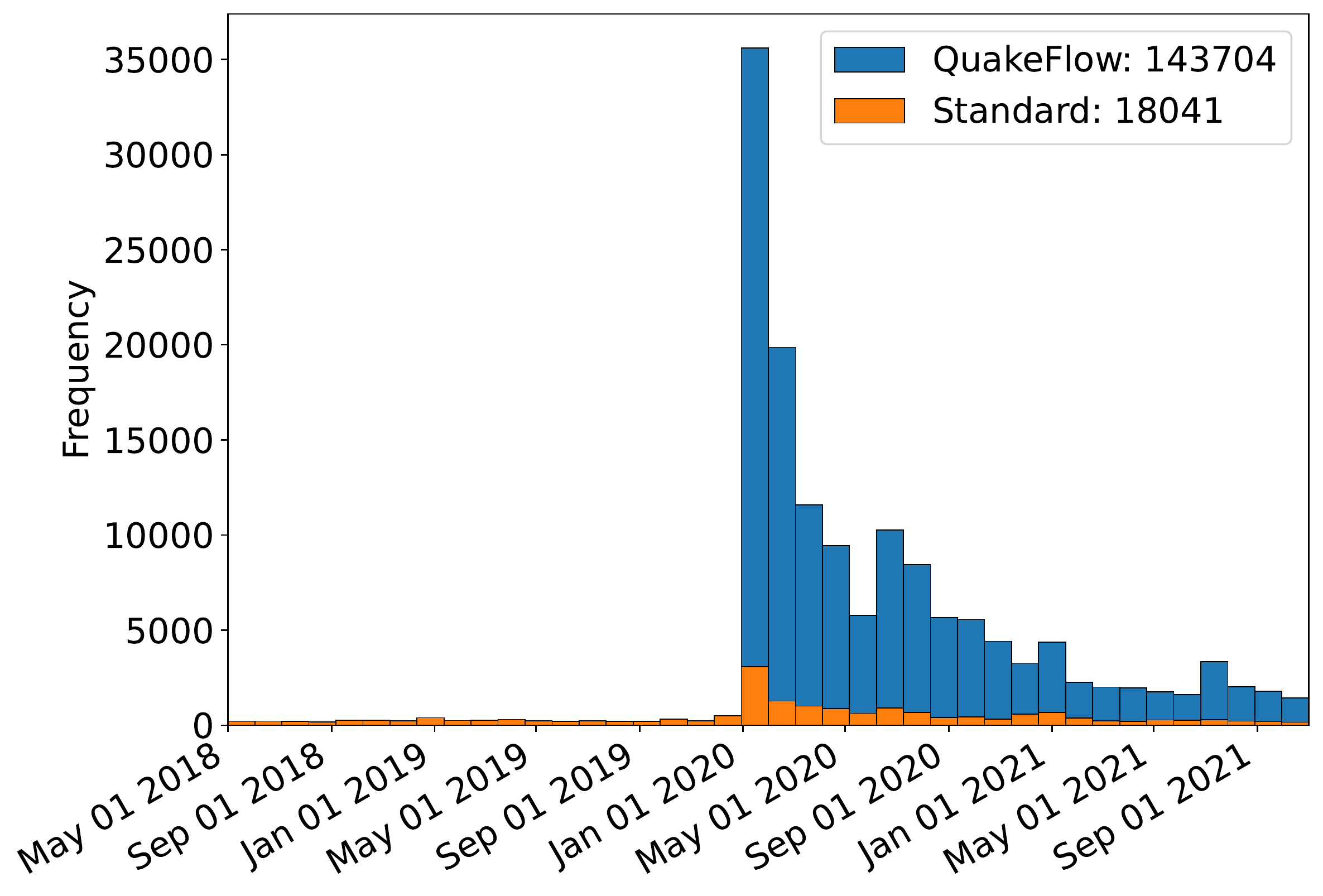}
    \caption{}
    \end{subfigure}
    \begin{subfigure}{0.48\textwidth}
    \includegraphics[width=\textwidth]{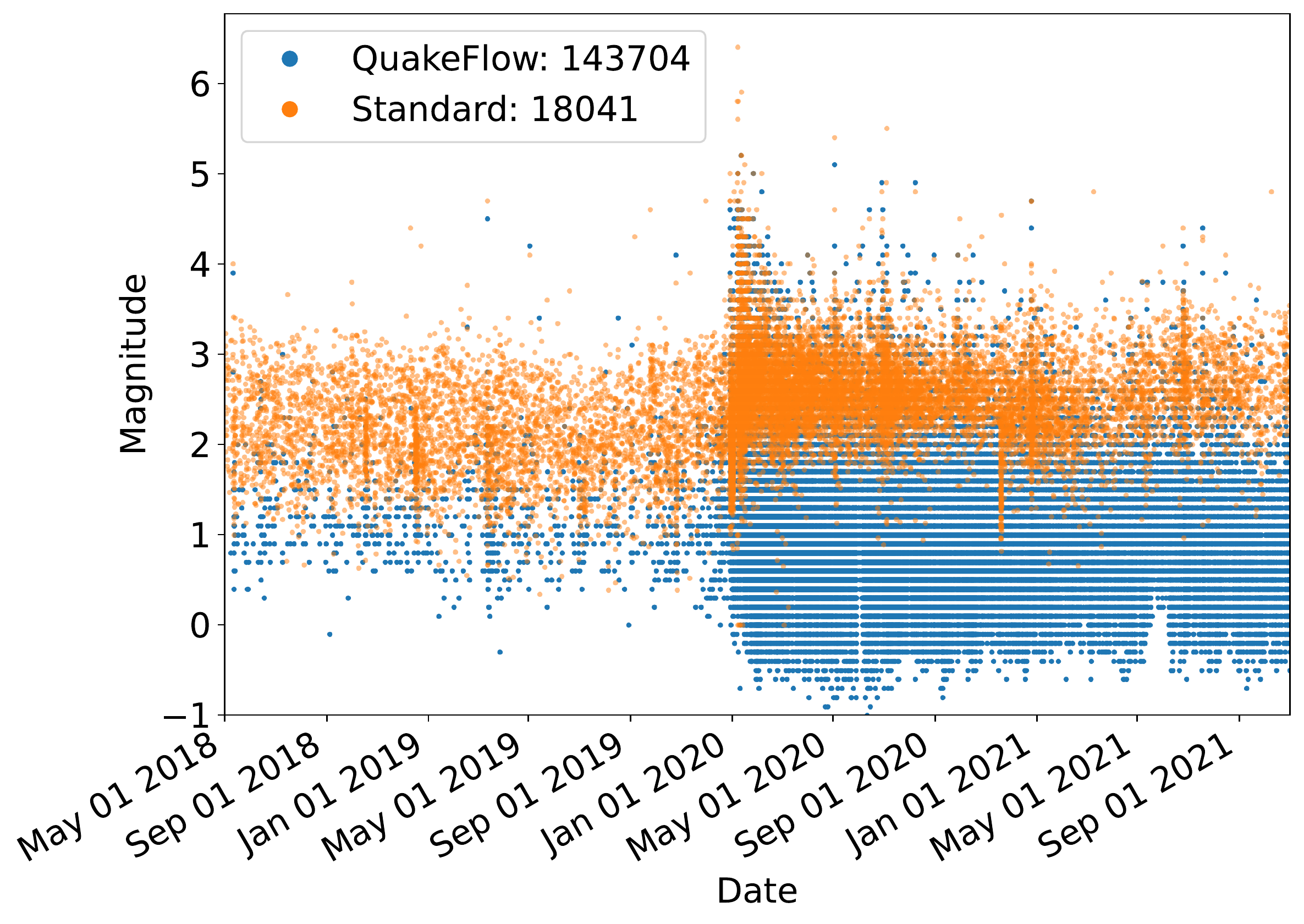}
    \caption{}
    \end{subfigure}
    \begin{subfigure}{0.44\textwidth}
    \includegraphics[width=\textwidth]{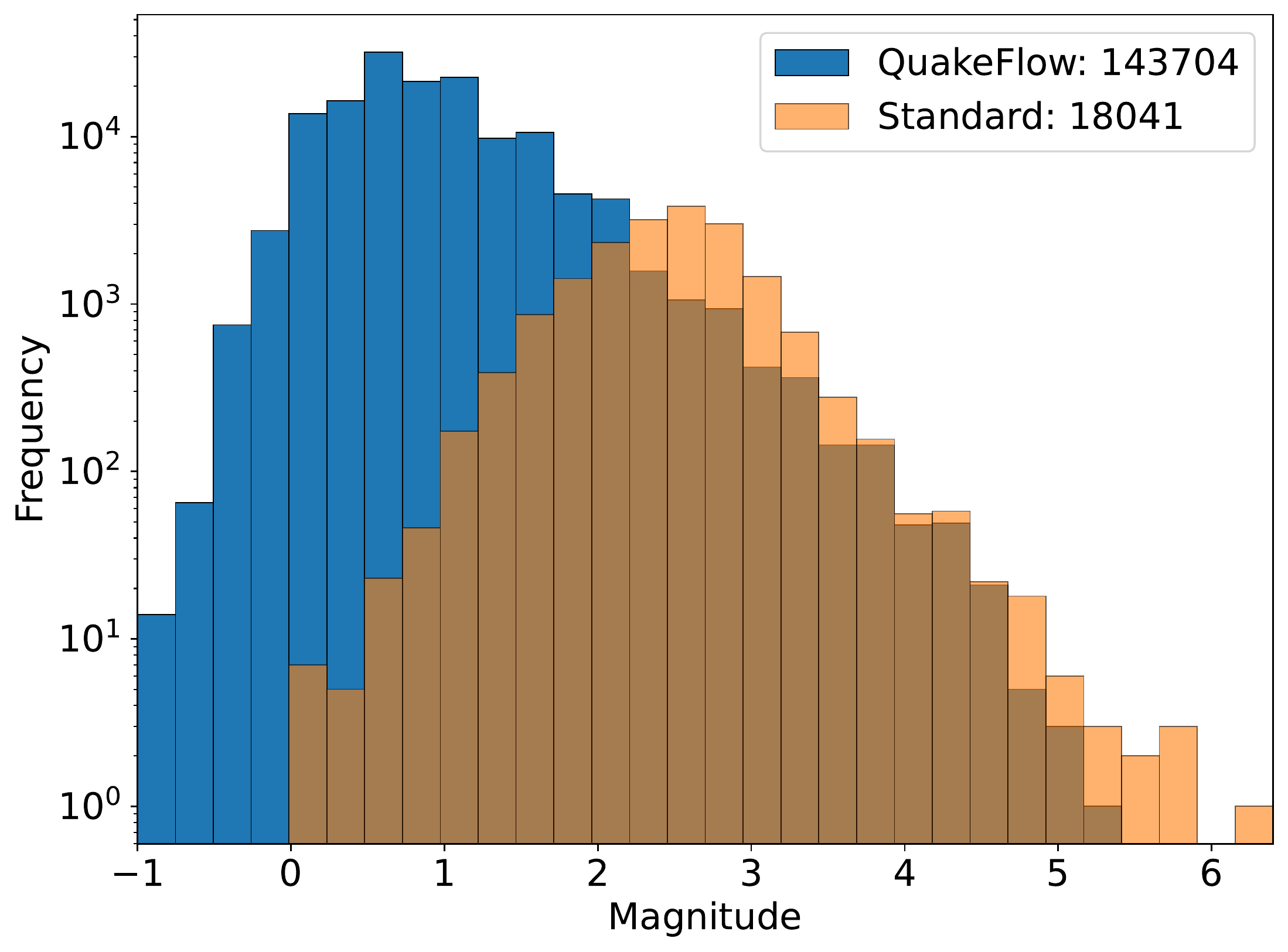}
    \caption{}
    \end{subfigure}
    \caption{Results for Puerto Rico: (a) seismic station locations; (b) earthquake frequency; (c) earthquake magnitude; (d) earthquake magnitude-frequency distribution. Blue indicates QuakeFlow results. Orange indicates the standard catalog. Note that earthquake magnitudes are estimated approximately during phase association using GaMMA.}
    \label{fig:puerto_rico}
\end{figure}

\begin{figure}
    \begin{subfigure}{0.48\textwidth}
    \includegraphics[width=\textwidth]{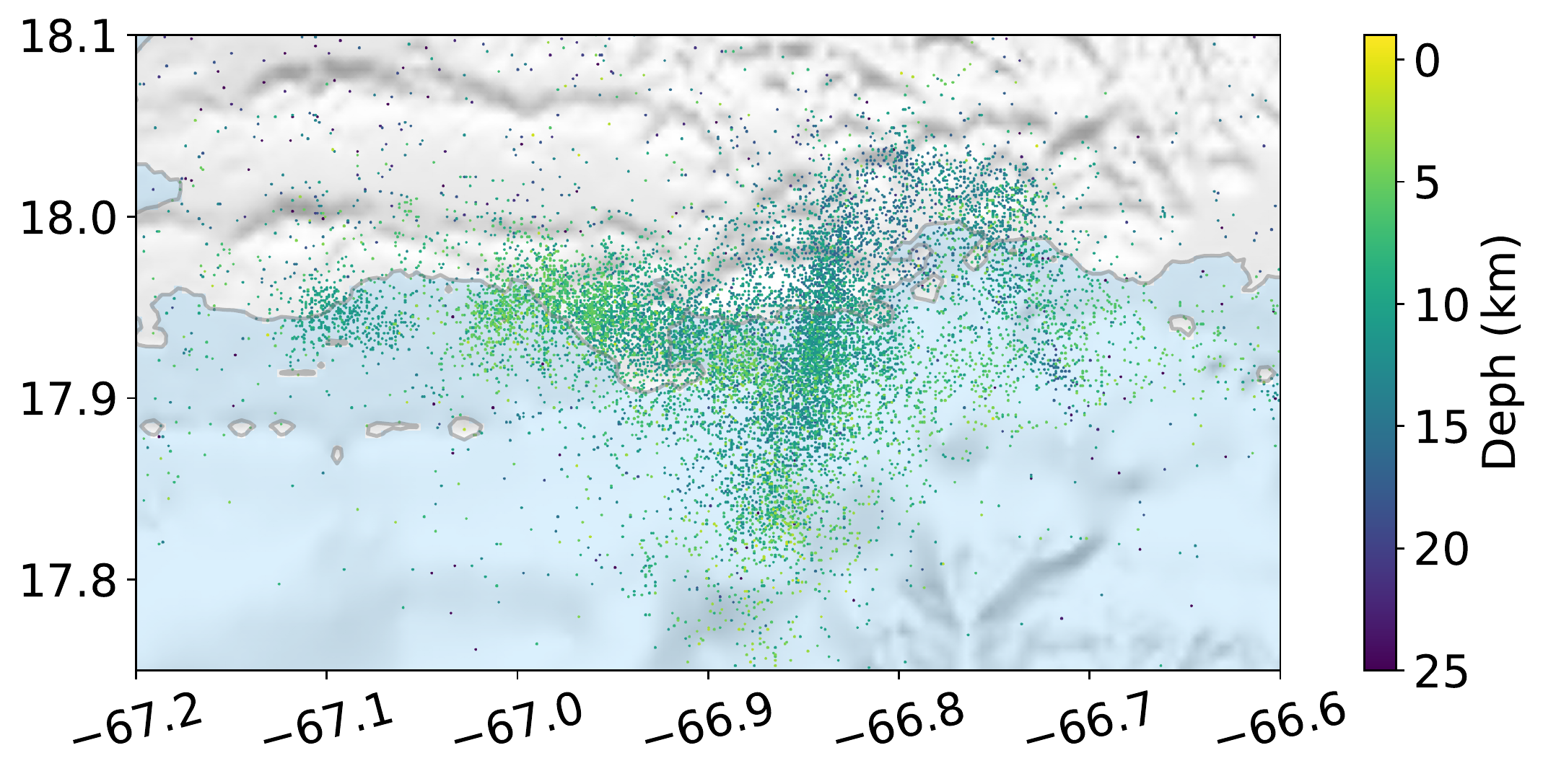}
    \caption{}
    \end{subfigure}
    \begin{subfigure}{0.48\textwidth}
    \includegraphics[width=\textwidth]{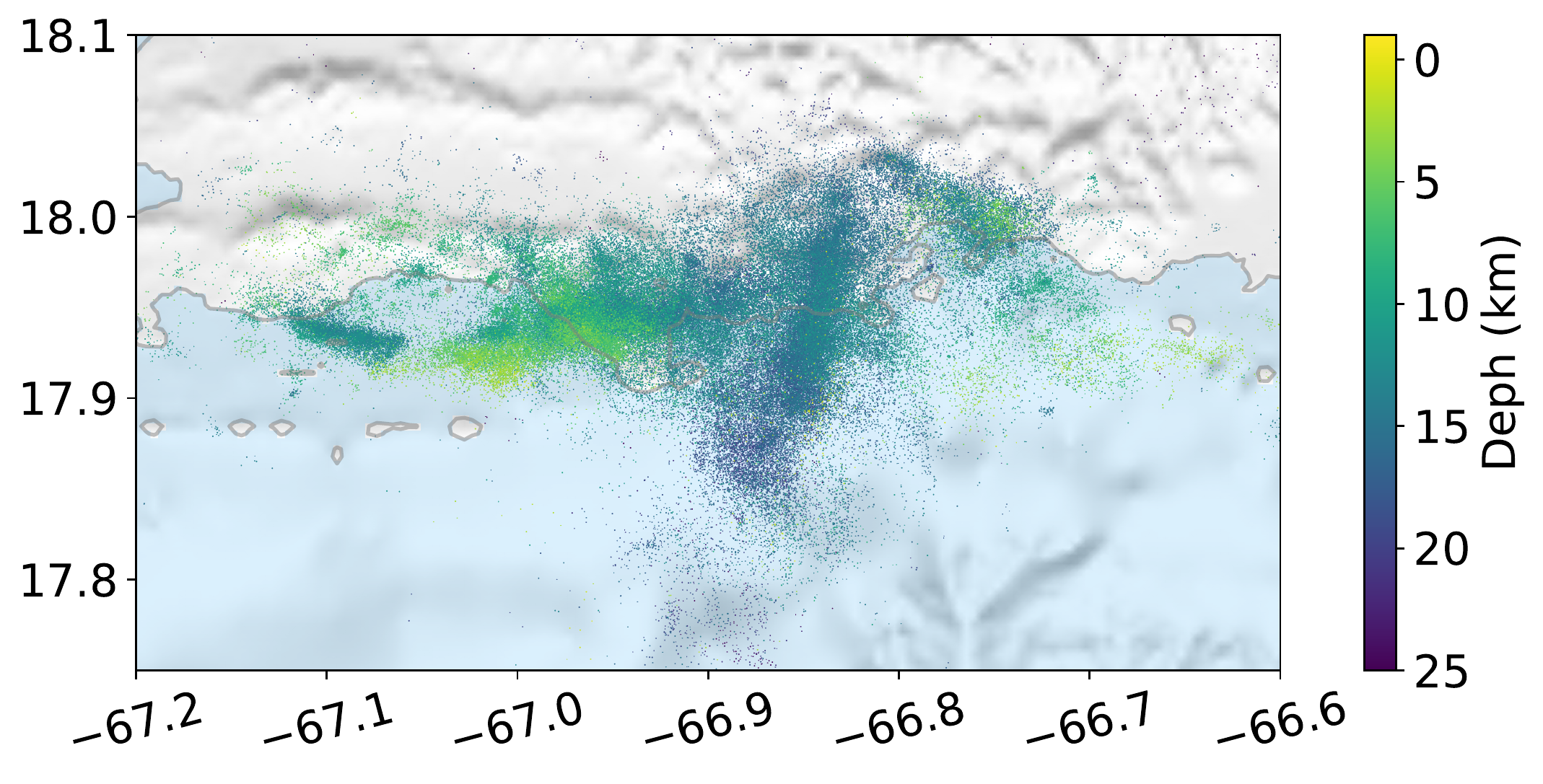}
    \caption{}
    \end{subfigure}
    \caption{Earthquake locations for Puerto Rico: (a) standard catalog; (b) QuakeFlow catalog. }
    \label{fig:puerto_rico_mapview}
\end{figure}

\begin{figure}
    \centering
    \begin{subfigure}{0.46\textwidth}
    \includegraphics[width=\textwidth]{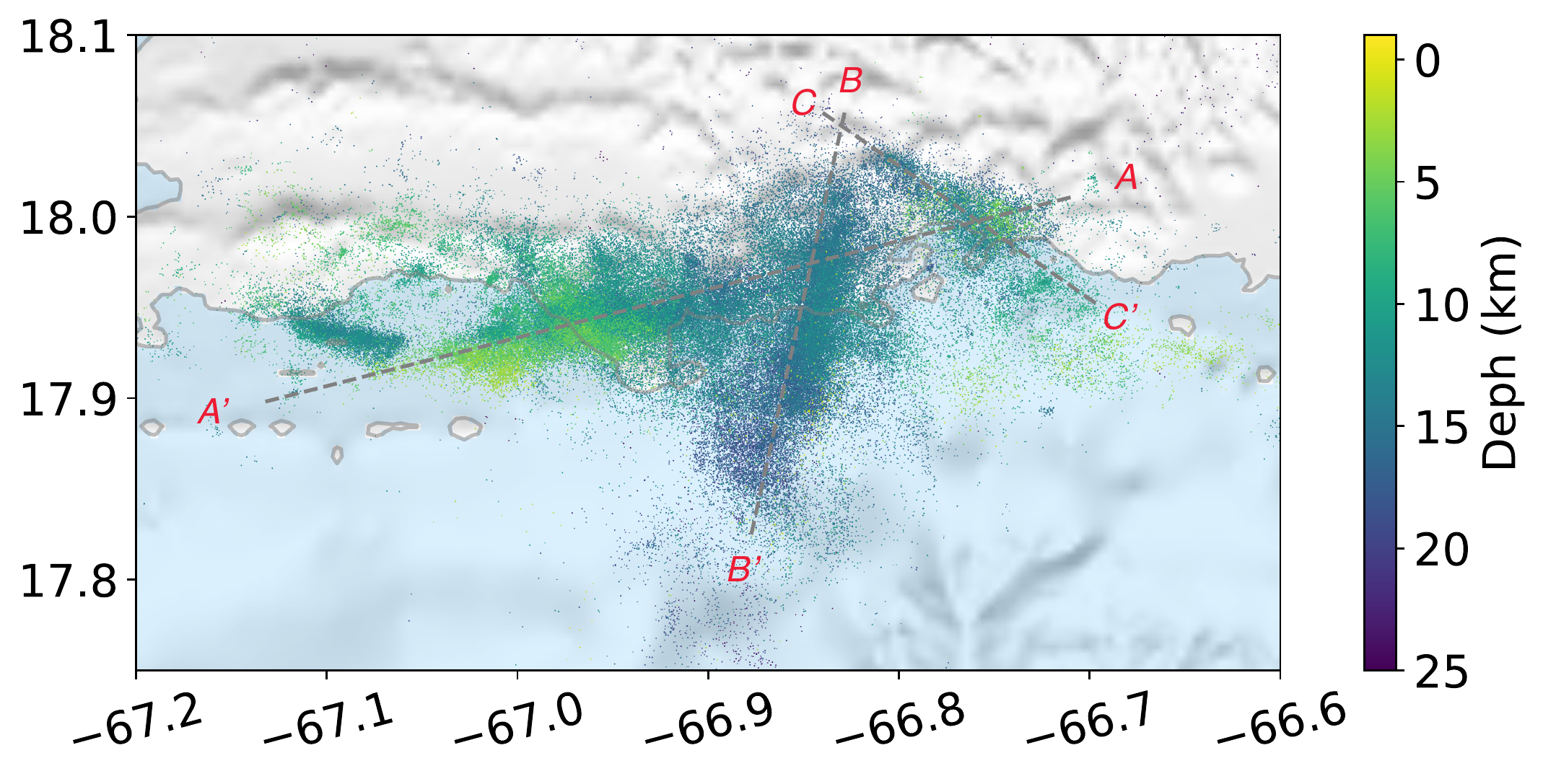}
    \caption{}
    \end{subfigure}
    \begin{subfigure}{0.53\textwidth}
    \includegraphics[width=\textwidth]{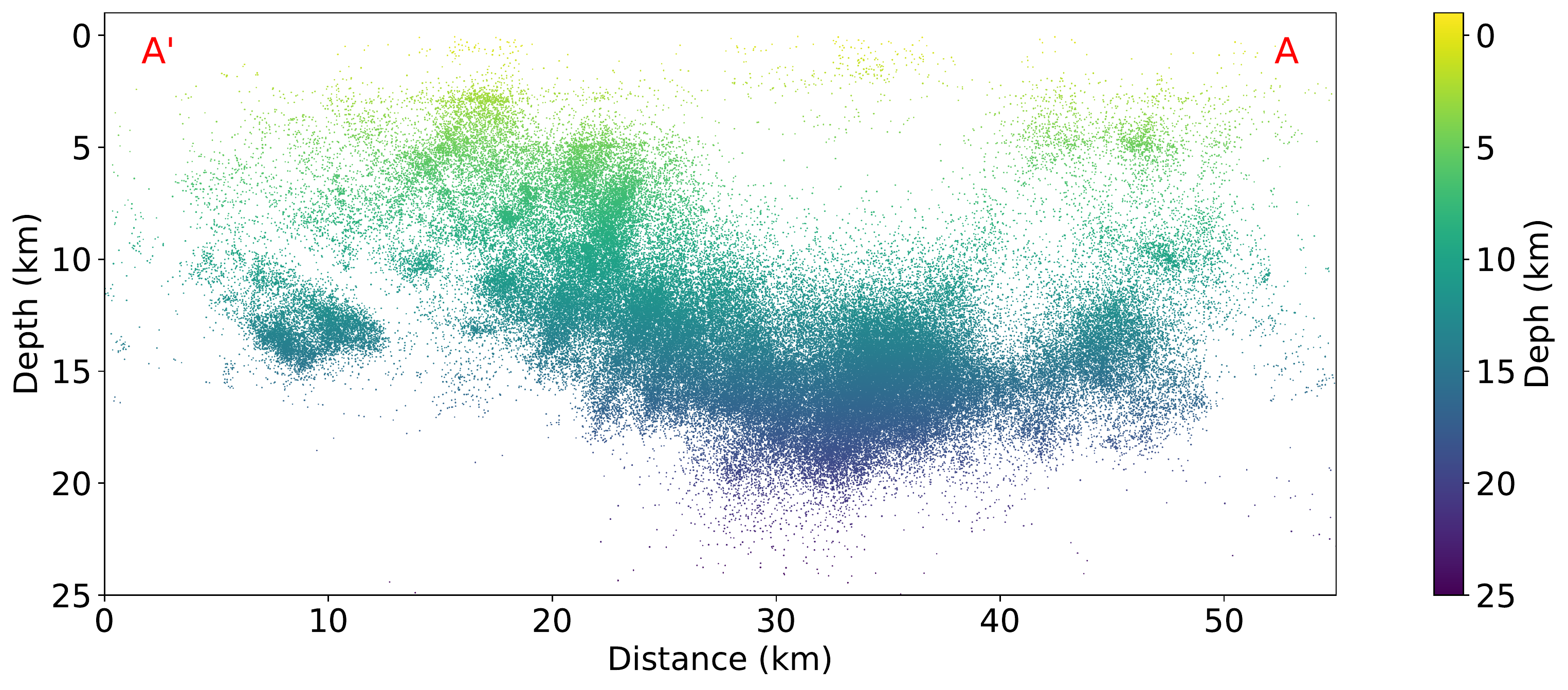}
    \caption{}
    \end{subfigure}
    \begin{subfigure}{0.50\textwidth}
    \includegraphics[width=\textwidth]{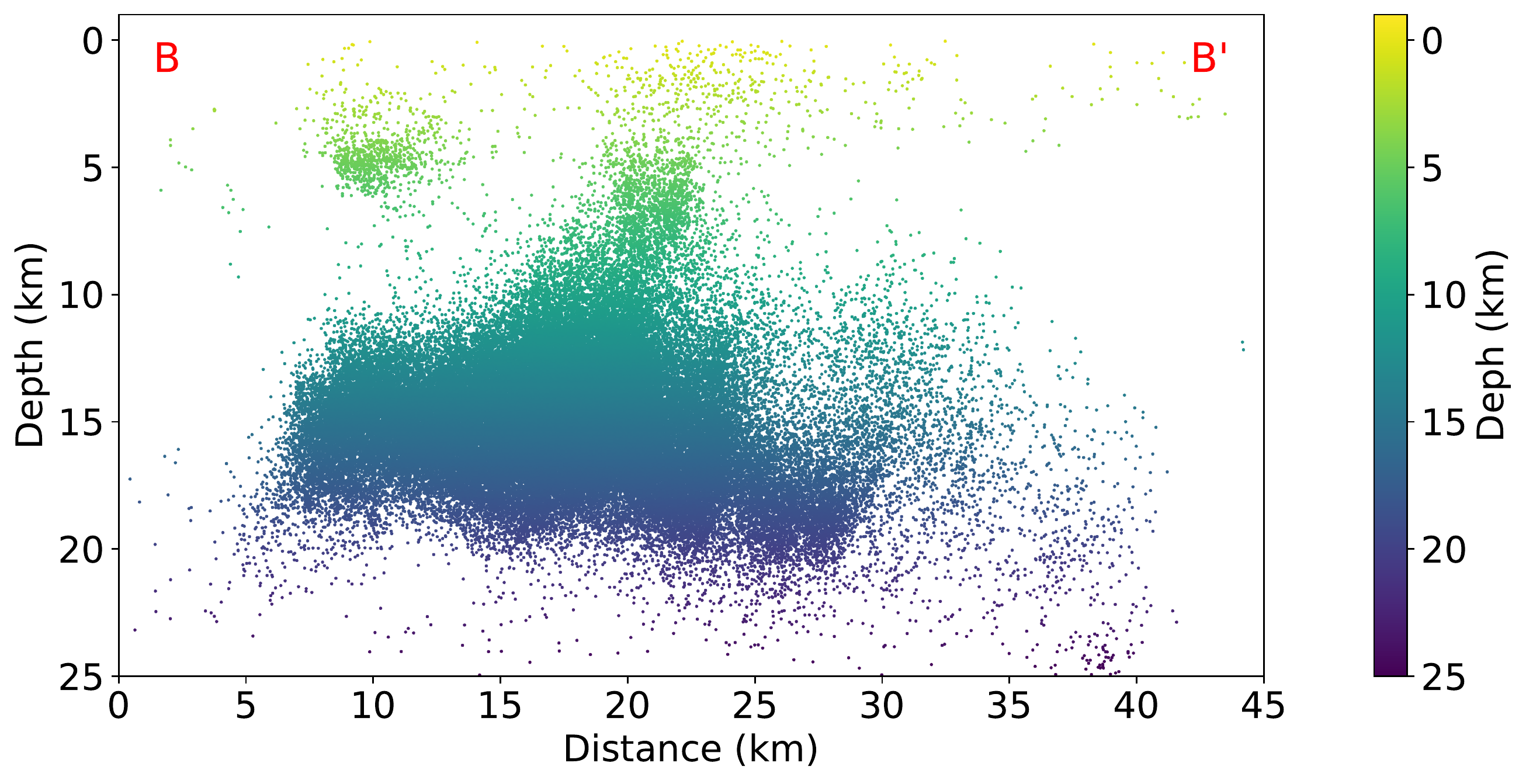}
    \caption{}
    \end{subfigure}
    \begin{subfigure}{0.49\textwidth}
    \includegraphics[width=\textwidth]{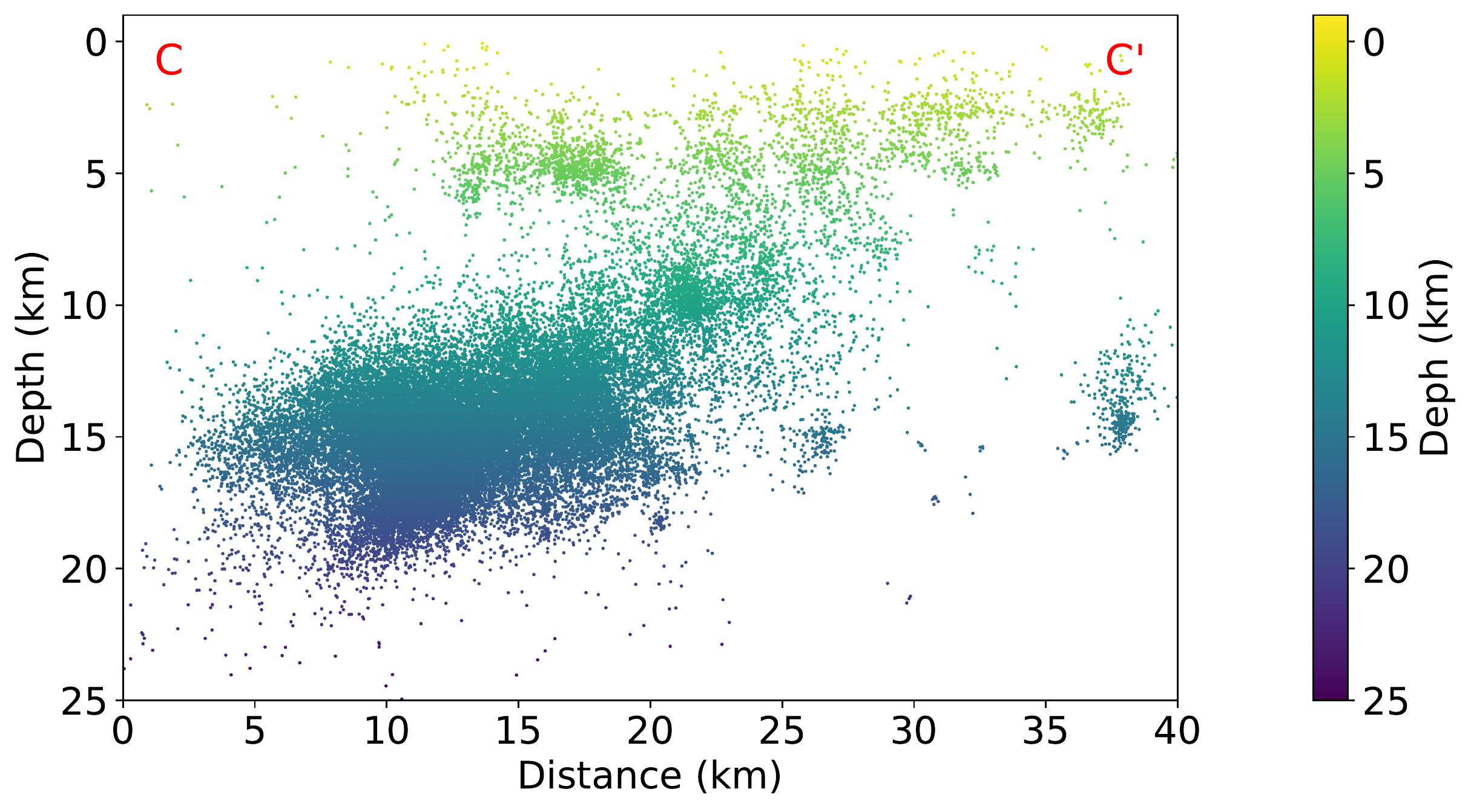}
    \caption{}
    \end{subfigure}
    \caption{Cross-sections of the earthquake catalog for Puerto Rico: (a) map view of three cross-sections; (b) A-A' cross-section; (b) B-B' cross-section; (c) C-C' cross-section. We plot events within 10 km of each of the cross-section lines in (a).}
    \label{fig:puerto_rico_slice}
\end{figure}

We focus on testing QuakeFlow's detection performance and processing speed. We applied QuakeFlow to three years of archived data from 2018-05-01 to 2021-05-01 using 70 stations within a region 65$^{\circ}$W - 68$^{\circ}$W and 17$^{\circ}$N - 19$^{\circ}$N (\Cref{fig:puerto_rico}a) for approximately 210 station-years of three-component continuous data from the Puerto Rico Seismic Network \citep{PR_FDSN} and the US Geological Survey Networks \citep{GS_FDSN}. We ran QuakeFlow on GCP with auto-scaling using a maximum of 60 computational nodes of machine type ``n2-standard-2" (2 vCPU and 8GB of memory). Downloading waveform data through the IRIS data center\footnote{\url{http://ds.iris.edu/ds/nodes/dmc/data/}} using ObsPy took approximately 3.5 hours, depending on internet conditions and data center server load. Picking P and S phase arrival times using PhaseNet took approximately 3 hours, and associating phases using GaMMA took an additional 30 minutes, or in other words, fast enough to run overnight. The total cost of this processing was around \$40 based on a price of \$0.07/hour per computational node\footnote{\url{https://cloud.google.com/compute/all-pricing}}. The earthquake detection results are shown in \Cref{fig:puerto_rico}b-d. Compared with the standard catalog generated by the Puerto Rico Seismic Network \citep{PR_FDSN}, QuakeFlow detected over an order of magnitude more small earthquakes, particularly during active aftershock periods after 2020-01-01. The exact number of earthquakes will vary depending on the hyperparameters, and there exists a trade-off between false positives and false negatives in both the standard catalog and the QuakeFlow catalog. Detailed comparison between these catalogs, and quantification of this trade-off, is an important direction for future research. The magnitudes in this workflow are based on simplified ground motion prediction equations \citep{zhu2022earthquake,picozzi2018rapid}. More accurate magnitudes could be determined by adding direct magnitude estimation to the QuakeFlow processing pipeline. The improved earthquake catalog provides important information to characterize aftershock activity, reveal detailed fault structure, and potentially improve aftershock forecasting (\Cref{fig:puerto_rico_mapview,,fig:puerto_rico_slice}).

\subsection{Earthquake detection in Hawaii}

\begin{figure}
    \centering
    \begin{subfigure}{0.4\textwidth}
    \includegraphics[width=\textwidth]{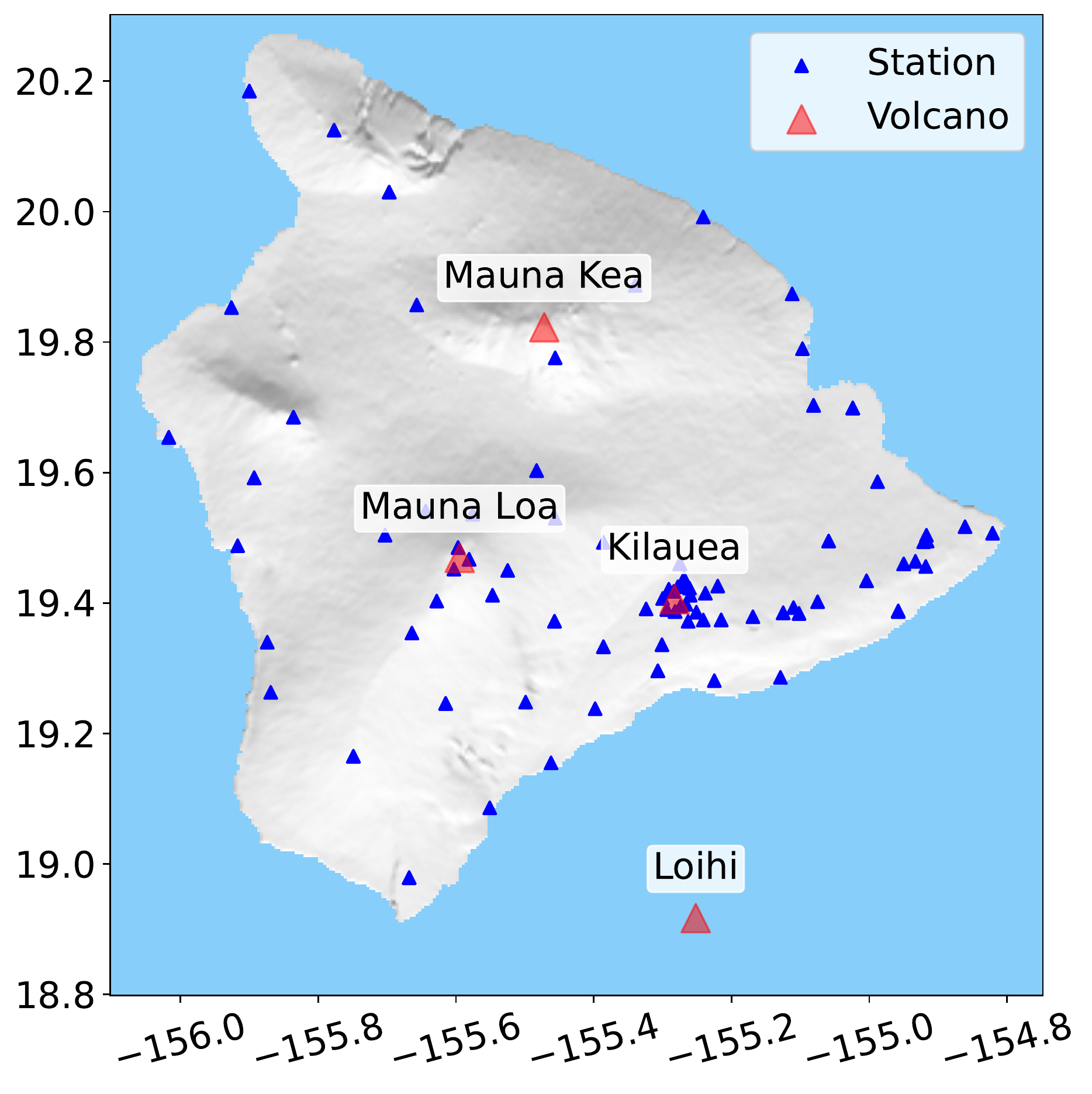}
    \caption{}
    \end{subfigure}
    \begin{subfigure}{0.48\textwidth}
    \includegraphics[width=\textwidth]{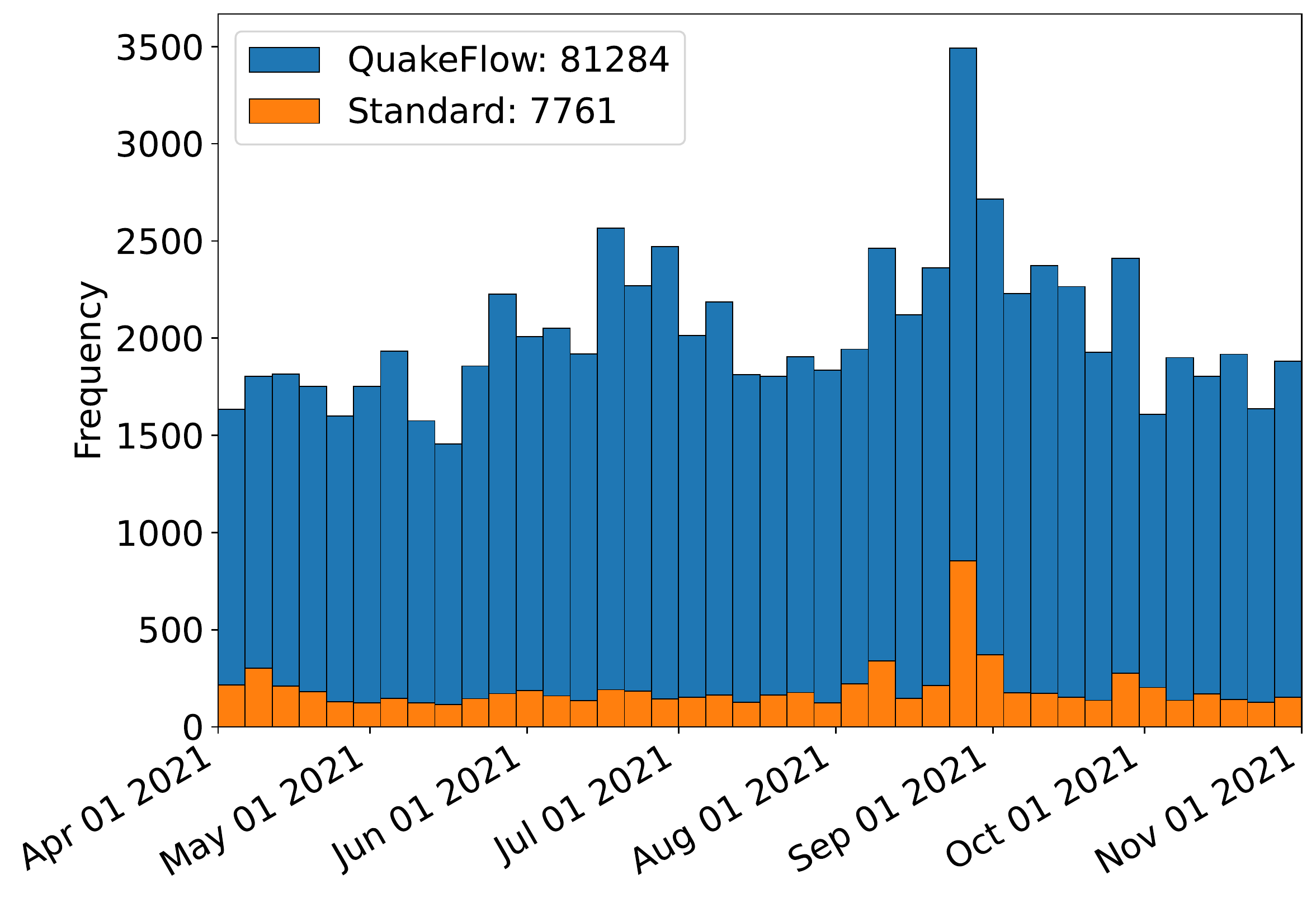}
    \caption{}
    \end{subfigure}
    \begin{subfigure}{0.48\textwidth}
    \includegraphics[width=\textwidth]{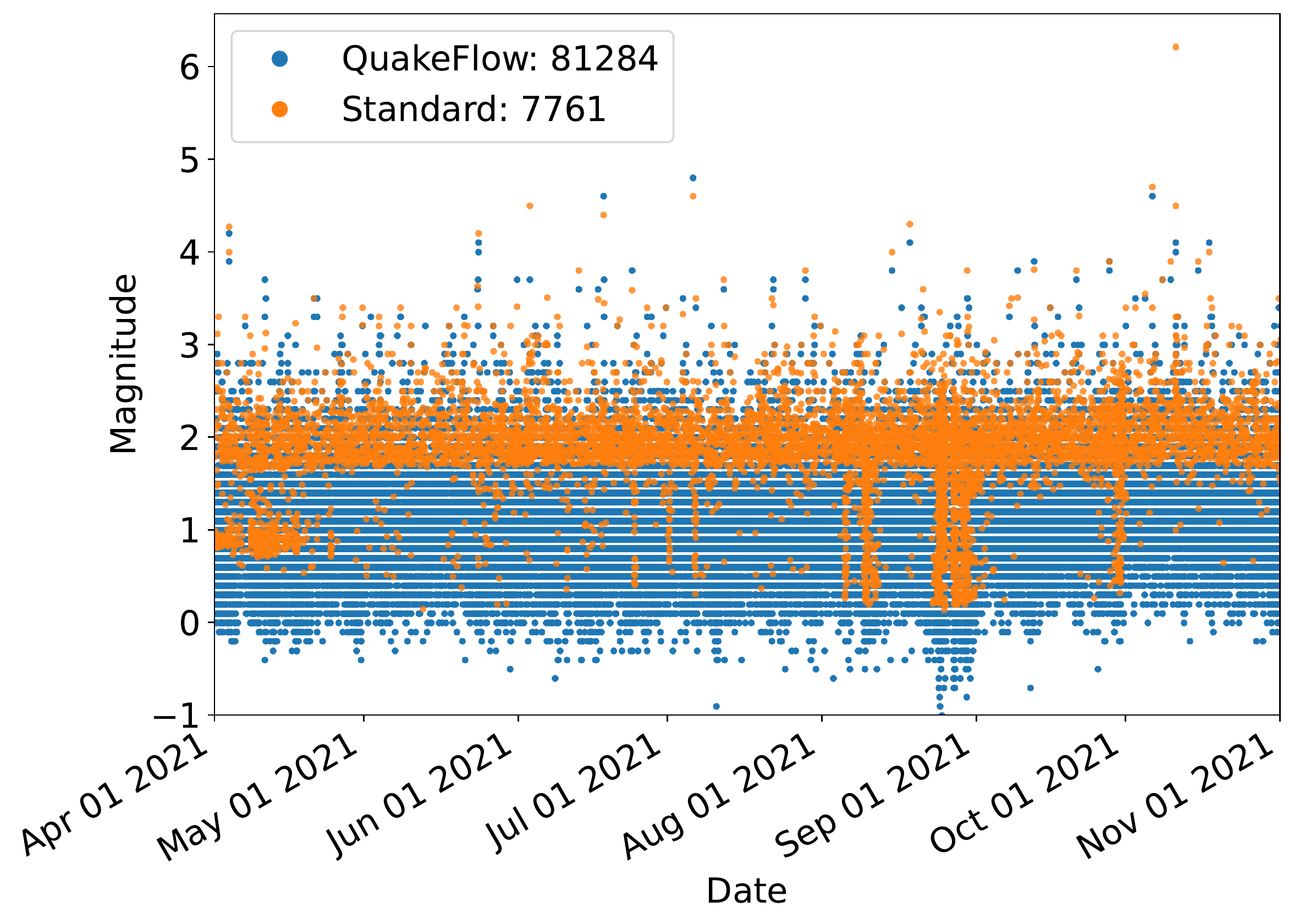}
    \caption{}
    \end{subfigure}
    \begin{subfigure}{0.44\textwidth}
    \includegraphics[width=\textwidth]{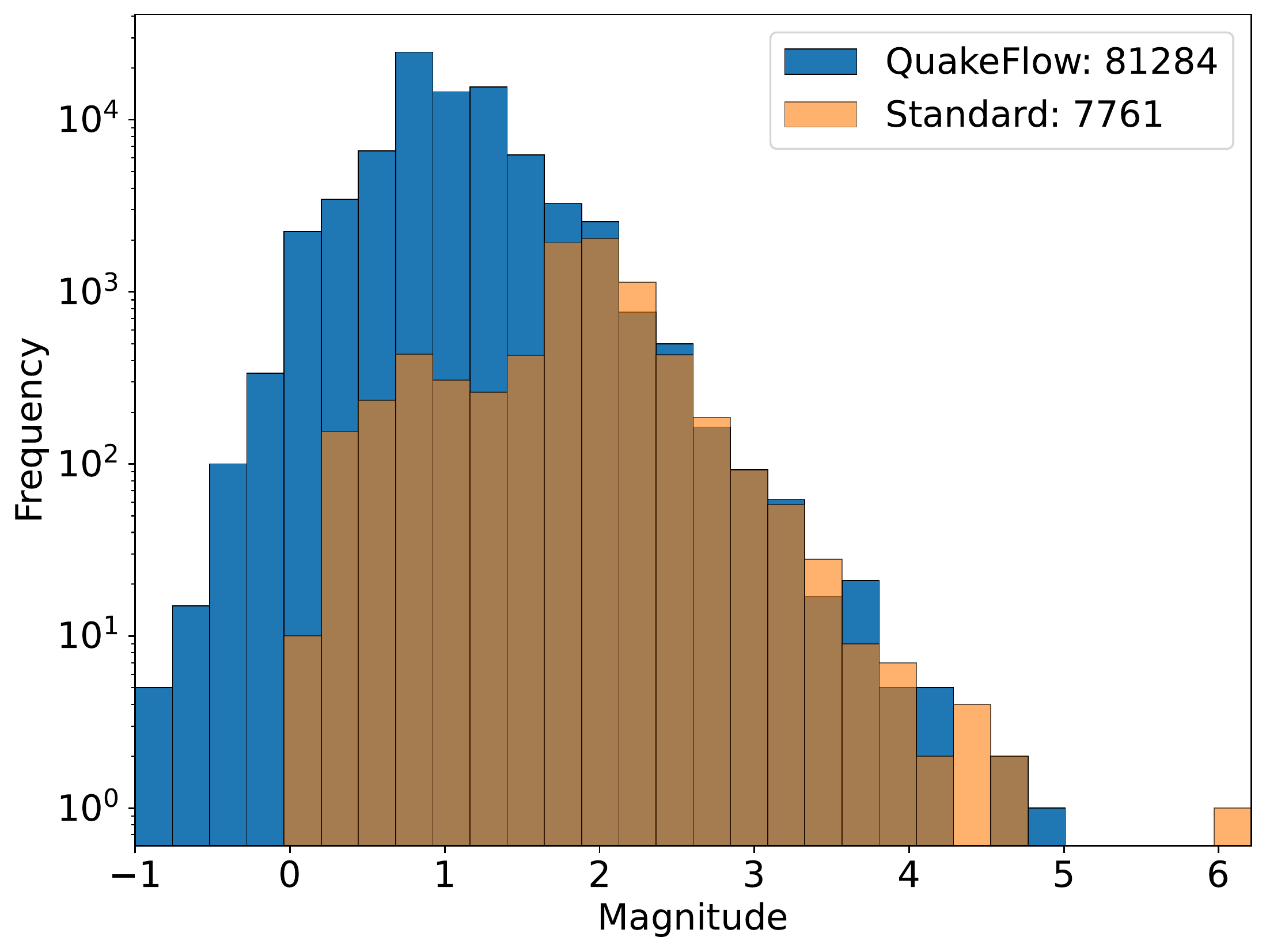}
    \caption{}
    \end{subfigure}
    \caption{Results for Hawaii: (a) seismic station locations; (b) earthquake frequency; (c) earthquake magnitude; (d) earthquake magnitude-frequency distribution. Blue indicates QuakeFlow results. Orange indicates the standard catalog. Note that  earthquake magnitudes are approximately estimated during phase association using GaMMA.}
    \label{fig:hawaii}
\end{figure}

We applied QuakeFlow to study volcanic earthquakes on the big island of Hawaii, which has seen a surge of eruptive activity, including the collapse of Kilauea Caldera (\Cref{fig:hawaii}a). Most volcanic earthquakes have small magnitudes, which make them an ideal target for QuakeFlow. The active magmatic system in Hawaii has prodigious seismicity, particularly during eruptions \citep{klein1987seismicity, matoza2021comprehensive}. Detecting and locating volcanic earthquakes can help illuminate the magma reservoirs and magmatic plumbing systems \citep{gillard1996highly, wech2015linking}. 
We retrieved seismic waveform data for 66 stations from the Hawaii Volcano Observatory Network (HVO) \citep{HV_FDSN}. With only a few hours of cloud computing, we obtain significantly improved resolution of seismicity over a broad depth range. As in Puerto Rico we detect over a factor of 10 more earthquakes than in the standard catalog reported by Hawaii Volcano Observatory Network over the same time period (\Cref{fig:hawaii}b-d), which can help illuminate the magmatic system.

\begin{figure}
    \begin{subfigure}{0.48\textwidth}
    \includegraphics[width=\textwidth]{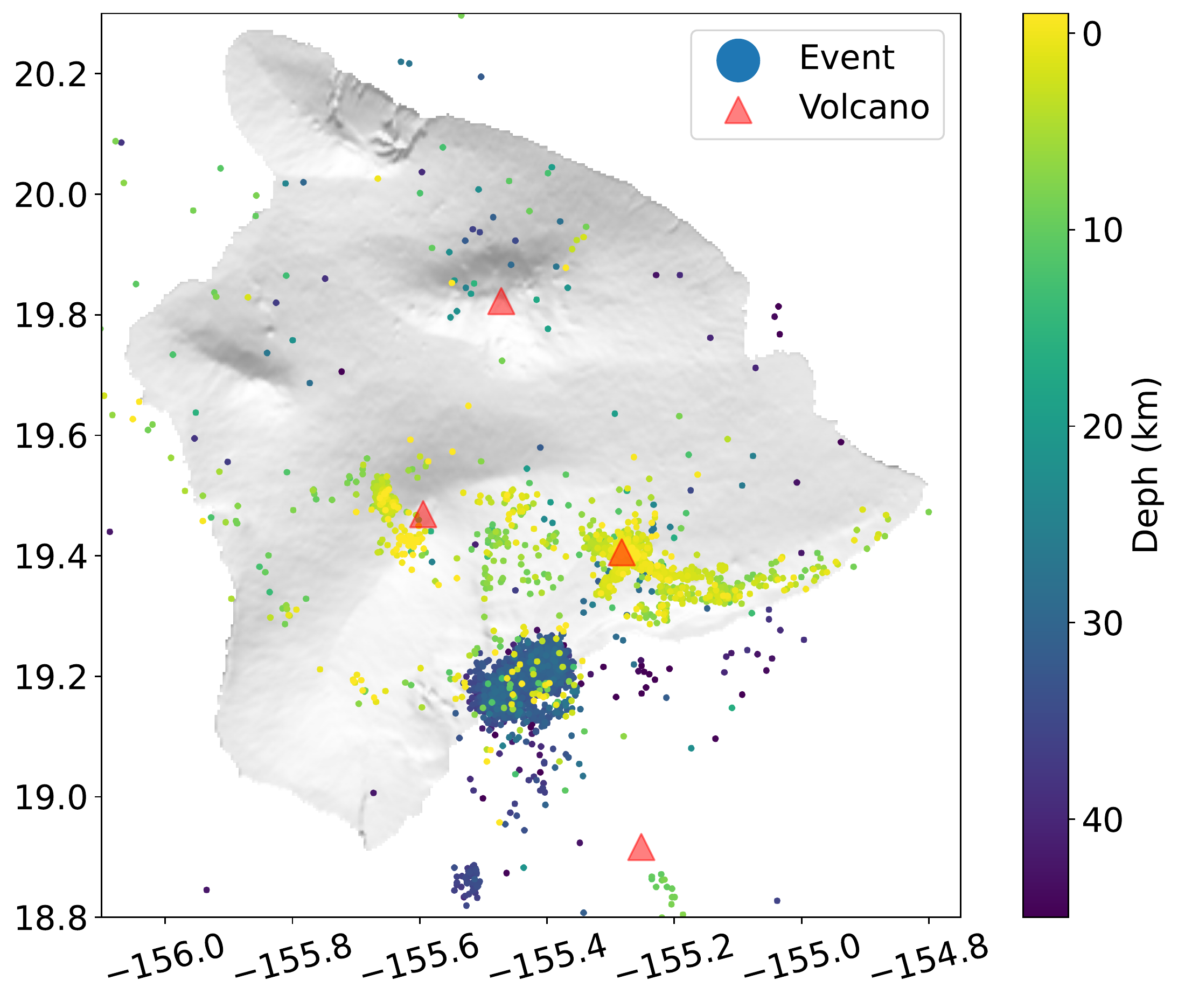}
    \caption{}
    \end{subfigure}
    \begin{subfigure}{0.48\textwidth}
    \includegraphics[width=\textwidth]{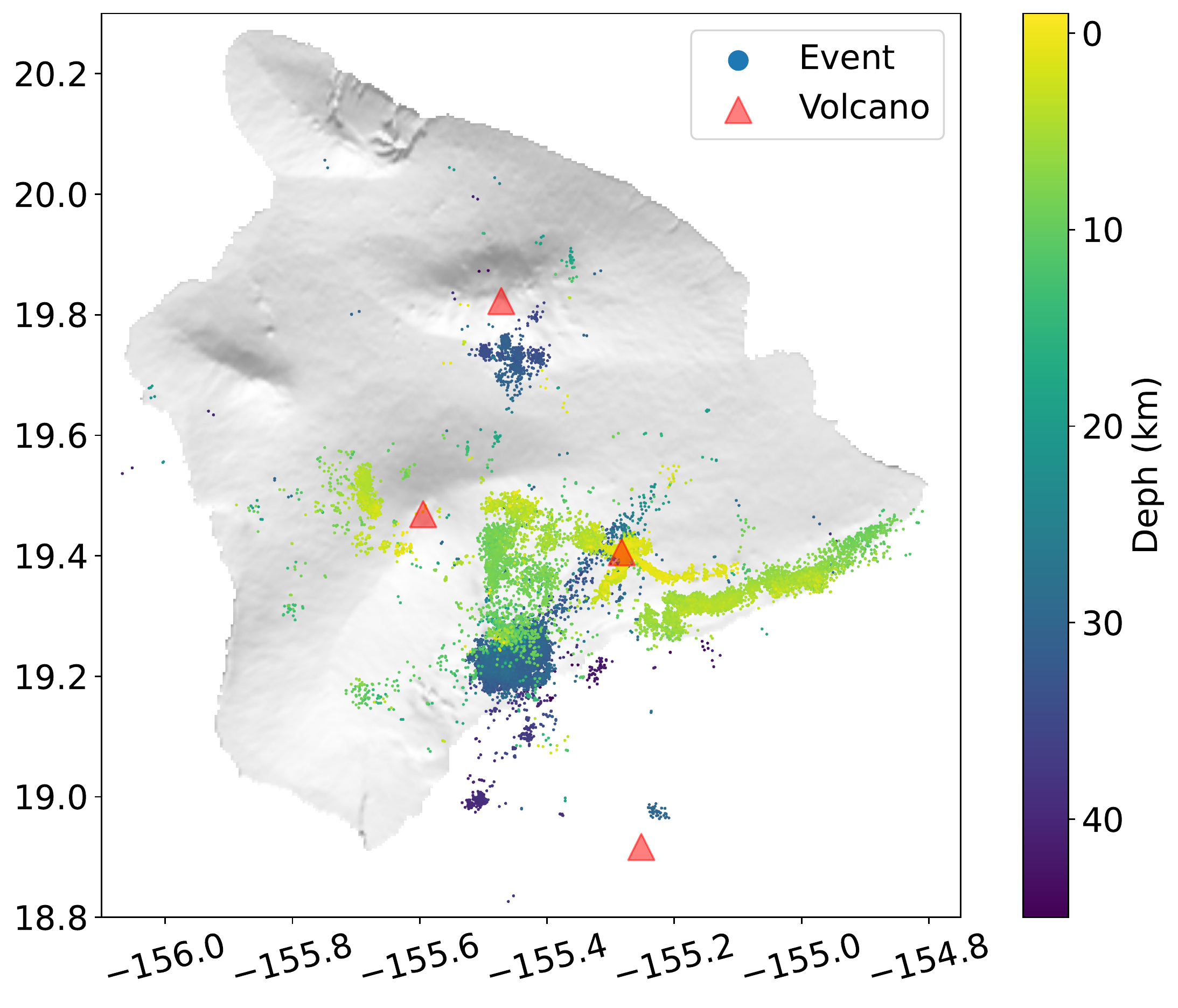}
    \caption{}
    \end{subfigure}
    \caption{Earthquake locations of Hawaii: (a) standard catalog; (b) QuakeFlow catalog.}
    \label{fig:hawaii_mapview}
\end{figure}

\begin{figure}
    \centering
    \begin{subfigure}{0.43\textwidth}
    \includegraphics[width=\textwidth]{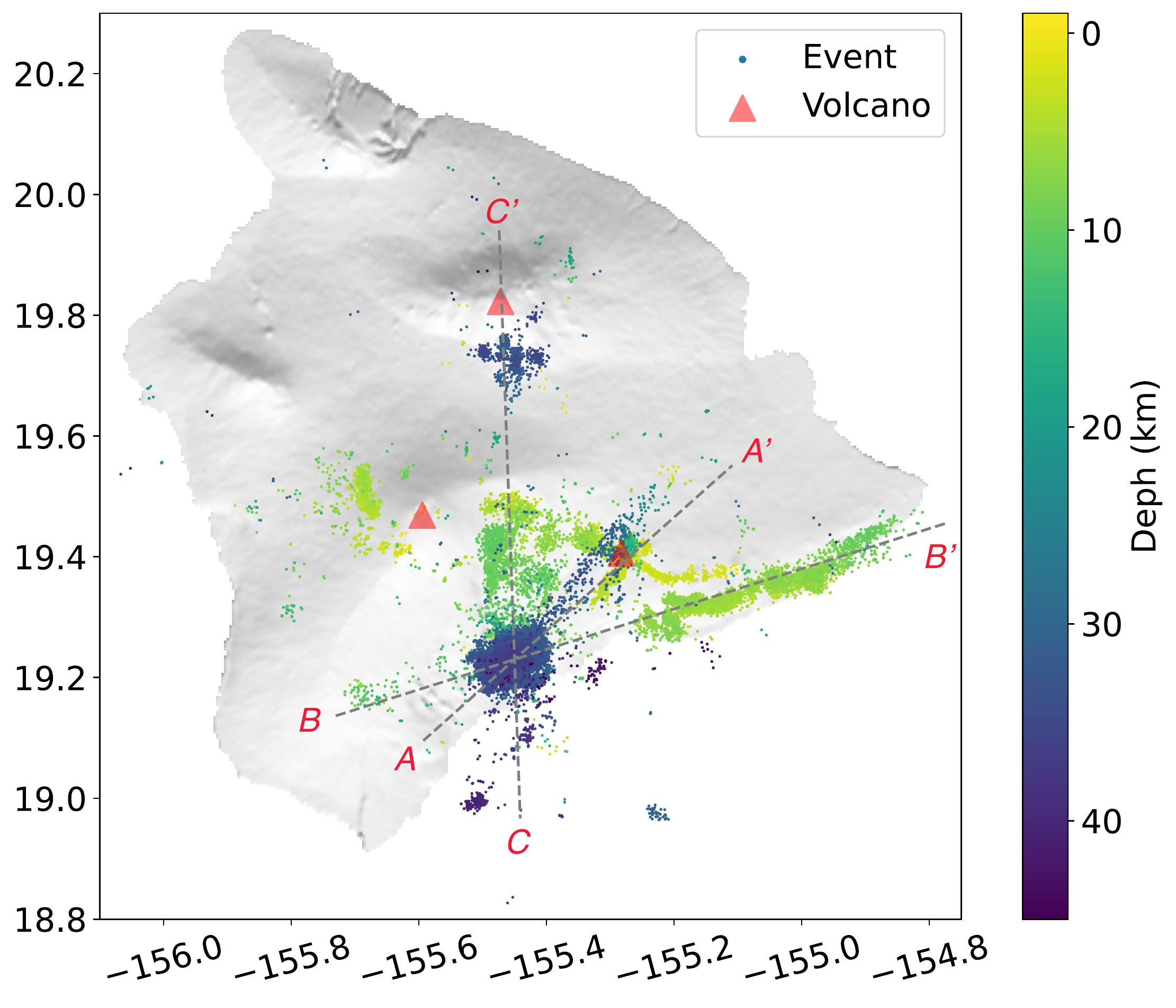}
    \caption{}
    \end{subfigure}
    \begin{subfigure}{0.45\textwidth}
    \includegraphics[width=\textwidth]{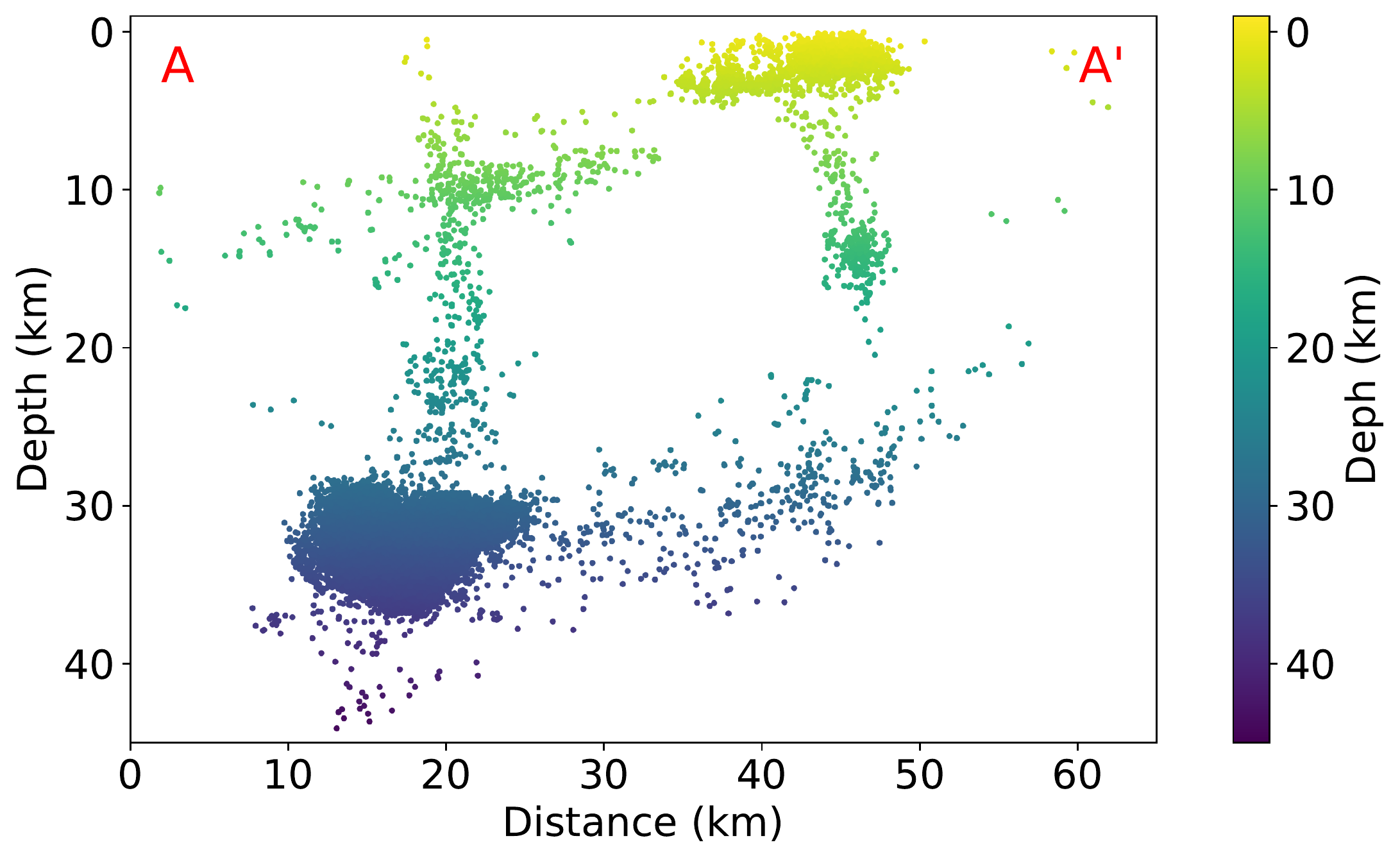}
    \caption{}
    \end{subfigure}
    \begin{subfigure}{0.50\textwidth}
    \includegraphics[width=\textwidth]{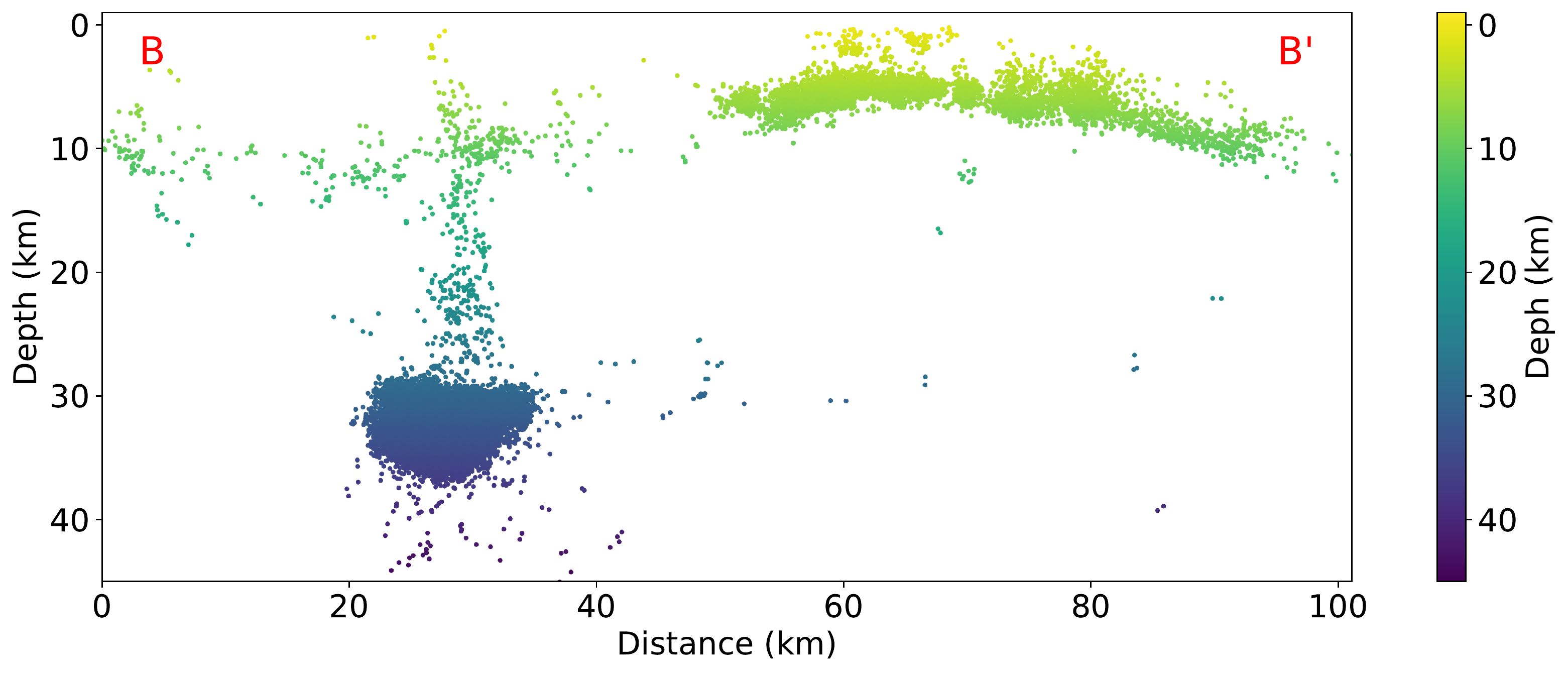}
    \caption{}
    \end{subfigure}
    \begin{subfigure}{0.49\textwidth}
    \includegraphics[width=\textwidth]{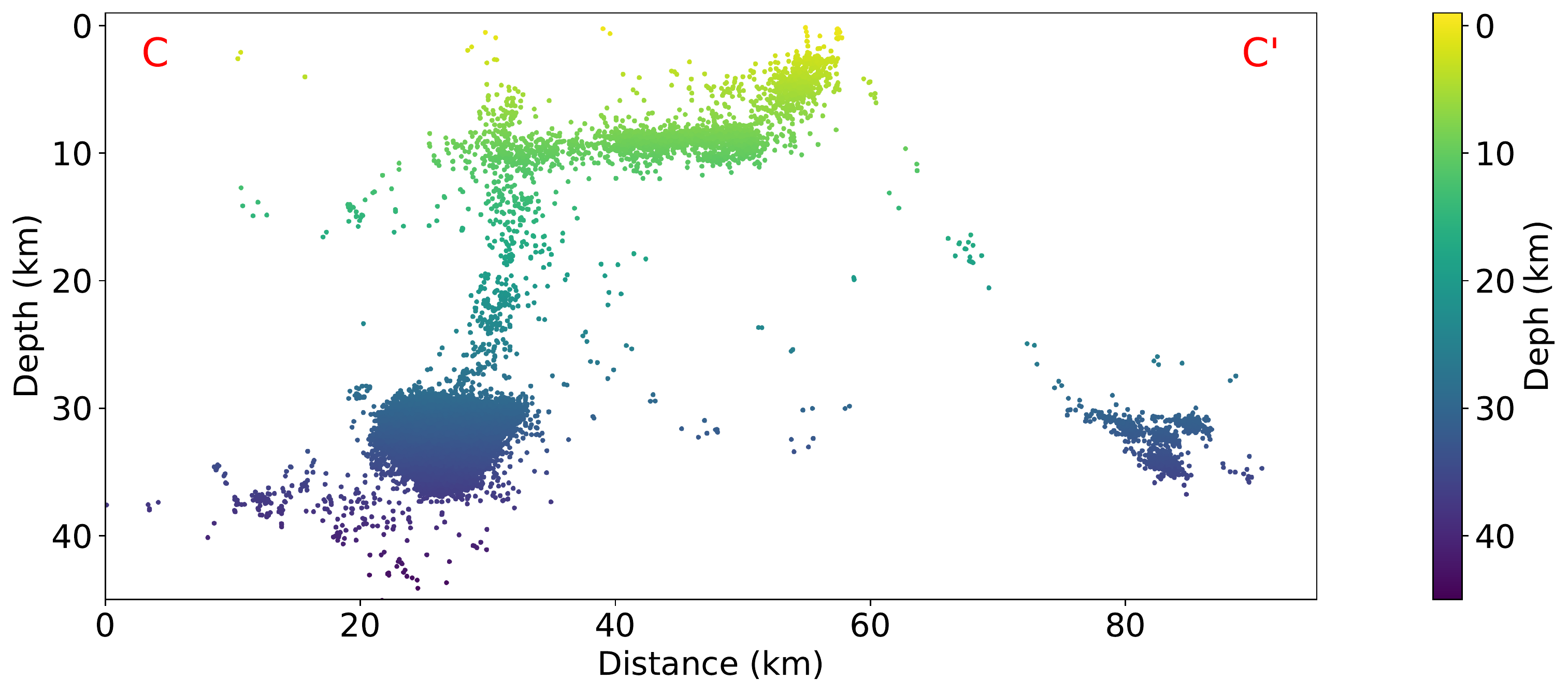}
    \caption{}
    \end{subfigure}
    \caption{Cross-sections of the earthquake catalog for Hawaii: (a) a map view of three cross-sections; (b) A-A' cross-section; (b) B-B' cross-section; (c) C-C' cross-section. Events within 5 km of the cross-section lines in (a) are plotted.}
    \label{fig:hawaii_slice}
\end{figure}

We find many deep events (below 30 km) in the Pāhala Mantle feature that showed a surge of activity since 2015 and is thought to be caused by the emplacement of new magma \citep{burgess2021ongoing} (\Cref{fig:hawaii_mapview,fig:hawaii_slice}). We also find both deep and shallow events forming lineations: one stretching to the northeast towards Kilauea (\Cref{fig:hawaii_slice}b), which is known as the mantle fault zone \citep{wolfe2003mantle} ; and another near the coastline (\Cref{fig:hawaii_slice}c), which corresponds to the decollement at about 10 km depth \citep{denlinger1995structure}.
More importantly, the improved catalog provides a clear picture of the connections between deep and shallow events, illuminating potential two magma transport paths from the deep Pāhala cluster to the Kilauea volcano (\Cref{fig:hawaii_slice}b): one migrating upward from the Pāhala cluster and the other migrating horizontally towards Kilauea along the mantle fault zone \citep{wright2006deep}.
There is also prodigious activity along the rift system, as expected, including additional shallow activity near the summit of Mauna Loa \citep{matoza2021comprehensive}, and deep events under Mauna Kea \citep{wech2020deep} as seen in cross-section C-C' (\Cref{fig:hawaii_slice}d). We note that all of these earthquakes are located with PhaseNet arrival time measurements. By adding cross-correlation-based arrival times, it would be possible to more clearly illuminate the full extent and geometric detail of active structures, to track temporal evolution, and to understand its relationship to eruptive activity.

\clearpage
\section{Discussion and Conclusions}

Machine learning - and especially deep learning - methods have developed rapidly in the last few years. Applications to several earthquake sequences (e.g., 2016-2017 the Central Apennines, Italy sequence \citep{tan2021machine}) have demonstrated that machine learning models trained on large labeled datasets can significantly outperform conventional approaches, resulting in an earthquake catalog with unprecedented spatial-temporal resolution. Applying these machine learning methods to revisit archived seismic data sets is a rewarding, but computationally challenging, task. Cloud computing addresses the computational challenge using almost unlimited computing nodes to parallelize seismic data mining workloads efficiently. As detailed here, we developed the QuakeFlow project to combine the impressive earthquake detection performance of machine learning algorithms with the powerful parallel processing capability of cloud computing to improve earthquake monitoring workflows. We built QuakeFlow based on the container-orchestration system, Kubernetes, and the Kubeflow project to run machine learning models in parallel on the cloud. QuakeFlow is made using containerized components, facilitating updates to deep learning/machine learning models current with the state-of-the-art. Quakeflow also facilitates benchmarking and comparison of the performance of competing algorithms by allowing swappable modules while holding data and other algorithms fixed. In its current implementation, QuakeFlow contains a deep neural network model for phase picking and a Gaussian mixture model for phase association to detect many more small earthquakes than conventional methods. Additional steps, such as denoising \citep{zhu2019seismic} or additional processing, such as precise earthquake location, magnitude, and focal mechanism determination, could be added to QuakeFlow to improve the catalog output. 

QuakeFlow explores a new approach to the earthquake monitoring workflow based on machine learning and cloud computing. We applied QuakeFlow to study both tectonic earthquakes in Puerto Rico and volcanic earthquakes in Hawaii. The results of these experiments demonstrate that cloud computing enables a flexible and efficient implementation of machine learning models for earthquake monitoring workflows. 
QuakeFlow runs on most cloud platforms with Kubernetes services to process huge amounts of seismic data in parallel with auto-scaling. It can be applied to many seismic networks and datasets to improve earthquake detection and reveal details of earthquake occurrence.

\section*{Acknowledgements}
We thank Miao Zhang, Yongsoo Park, and Ian McBrearty for helpful discussions. 
The facilities of IRIS Data Services, and specifically the IRIS Data Management Center, were used for access to waveforms, related metadata, and/or derived products used in this study. 
This work was supported by AFRL under contract number FA9453-19-C-0073.

\section*{Data Availability}
The seismic waveforms and earthquake catalogs used in this study come from the Puerto Rico Seismic Network, the US Geological Survey Networks, and the Hawaiian Volcano Observatory Network. The \href{https://github.com/wayneweiqiang/QuakeFlow}{QuakeFlow} codebase is accessible at \href{https://www.doi.org/10.5281/zenodo.7023970}{DOI: 10.5281/zenodo.7023970}

\bibliographystyle{apacite}
\bibliography{reference}

\end{document}